\documentclass[fleqn,usenatbib,usedcolumn]{mnras}
  \usepackage{amsmath}
  \usepackage{upgreek}
   \usepackage{hyperref}
   \usepackage[british]{babel}             % British English hyphenation
   \usepackage{txfonts}                    % Reasonable fonts
   \let\la=\lesssim % for less than similar from txfonts, not \la from mnras.cls
   \usepackage{graphicx}                   % Including figures

   \hypersetup{pdfauthor={Jos\'e Fonseca, Marta B. Silva, M\'ario G. Santos and Asantha Cooray},
               pdftitle={Cosmology with intensity mapping techniques using atomic and molecular lines},
               pdfkeywords={cosmology: large-scale structure of the Universe, miscellaneous},
               bookmarksnumbered=true}
   \hypersetup{colorlinks=true,
            linkcolor=blue,
            citecolor=blue,
            filecolor=blue,
            urlcolor=blue}

\newcommand{\be}{\begin{equation}}
\newcommand{\ee}{\end{equation}}
\newcommand{\bea}{\begin{eqnarray}}
\newcommand{\eea}{\end{eqnarray}}
\newcommand{\bfig}{\begin{figure}}
\newcommand{\efig}{\end{figure}}
\newcommand{\nn}{\nonumber}

\def\l{\left(}
\def\r{\right)}

\def\D{\Delta}
\def\la{\lambda}
\def\hmf{\frac{dn}{dM}}
\def\a{{\alpha}}
\def\d{\mathrm{d}}
\def\z{\l z\r}
\def\lya{{\rm Ly\alpha}}
\def\uv{{\rm UV}}
\def\ha{{\rm H\alpha}}

\title[Cosmology with line IM]{Cosmology with intensity mapping techniques using atomic and molecular lines}
\author[Fonseca et al.]{Jos\'e Fonseca$^1$\thanks{josecarlos.s.fonseca@gmail.com}, Marta B. Silva$^2$, M\'ario G. Santos$^{1,3,4}$ \& Asantha Cooray$^5$\\
$^1$ Department of Physics and Astronomy, University of the Western Cape, Cape Town 7535, South Africa\\
$^2$ Kapteyn Astronomical Institute, University of Groningen, Landleven 12, 9747AD Groningen, the Netherlands\\
$^3$ SKA SA, The Park, Park Road, Cape Town 7405, South Africa\\
$^4$ CENTRA, Instituto Superior T\'ecnico, Universidade de Lisboa, 1049-001 Lisboa, Portugal\\ 
$^5$ Department of Physics \& Astronomy, University of California, Irvine, CA 92697\\
}
\date{}

\pagerange{\pageref{firstpage}--\pageref{lastpage}}

\begin{document}

\label{firstpage}

\maketitle

\begin{abstract}
{We present a systematic study of the intensity mapping (IM) technique using updated models for the different emission lines from galaxies. We identify which ones are more promising for cosmological studies of the post reionization epoch. We consider the emission of $\lya$, $\ha$, H$\beta$, optical and infrared oxygen lines, nitrogen lines, CII and the CO rotational lines. We show that $\lya$, $\ha$, OII, CII and the lowest rotational CO lines are the best candidates to be used as IM probes. These lines form a complementary set of probes of the galaxies emission spectra. We then use reasonable experimental setups from current, planned or proposed experiments to assess the detectability of the power spectrum of each emission line. Intensity mapping of  $\lya$ emission from $z=2$ to 3 will be possible in the near future with HETDEX, while far-infrared lines require new dedicated experiments. We also show that the proposed SPHEREx satellite can use OII and $\ha$ IM to study the large-scale distribution of matter in intermediate redshifts of 1 to 4. We find that submilimeter experiments with bolometers can have similar performances at intermediate redshifts using CII and CO(3-2).}
\end{abstract}

% Don't make up new ones.
\begin{keywords}
cosmology: large-scale structure of the Universe, miscellaneous.
\end{keywords}

%%%%%%%%%%%%%%%%%%%%%%%%%%%%%%%%%%%%%%%%%%%%%%%%%%

%%%%%%%%%%%%%%%%% BODY OF PAPER %%%%%%%%%%%%%%%%%%

% The MNRAS class isn't designed to include a table of contents, but for this document one is useful.
% I therefore have to do some kludging to make it work without masses of blank space.
%\begingroup
%\let\clearpage\relax
%\tableofcontents
%\endgroup
%\newpage

\section{Introduction}

Measurements of the 3-dimensional (3D) large-scale structure of the Universe across cosmic time promise to bring exquisite constraints on cosmology, from the nature of dark energy or the mass of neutrinos, to primordial non-Gaussianity and tests of General Relativity (GR) on large scales. Most of these surveys are based on imaging a large number of galaxies at optical or near-infrared wavelengths, combined with redshift information to provide 3D positions of the galaxies [BOSS (SDSS-III) \citep{2009astro2010S.314S}, DES \citep{Flaugher:2004vg}, eBOSS \citep{Dawson:2015wdb}, DESI \citep{Levi:2013gra}, 4MOST \citep{deJong:2012nj}, LSST \citep{Abate:2012za,Bacon:2015dqe}, WFIRST \citep{Spergel:2015sza}, and the Euclid satellite \citep{Laureijs:2011gra,Amendola:2012ys}]. These observations at optical and infrared wavelengths will be limited to galaxy samples between $z=0.3$ to 2, and in some cases, redshifts will be determined only by the photometric data.

Instead of counting galaxies, the intensity mapping (IM) technique uses the total observed intensity from any given pixel. For a reasonably large 3D pixel (with a given angular and frequency/redshift resolution), also referred to as \emph{voxel}, we expect it to contain several galaxies. The intensity in each pixel will thus be the integrated emission from all these galaxies. This should then provide a higher signal-to-noise compared to standard galaxy ``threshold" surveys. Moreover, since most cosmological applications rely on probing large scales, the use of these large pixels will not affect the cosmological constraints. By not needing galaxy detections, the requirements on the telescope/survey will be much less demanding. However, since we are no longer relying on ``clean" galaxy counts, we need to be much more careful with other contaminants of the observed intensity.

In order to have redshift information, the measured intensity should originate from specific emission lines. The underlying idea is that the amplitude of this intensity will be related to the number of galaxies in the 3D pixel emitting the target line. The fluctuations in the intensity across the map should then be proportional to the underlying dark matter fluctuations. Several lines can in principle be used for such surveys. In particular, a significant focus has been given in recent years to the HI 21cm line \citep{Battye:2004re, 2013MNRAS.434L..46S, Chang:2007xk, 2008PhRvL.100p1301L,% 2008MNRAS.383..606W, 2008MNRAS.383.1195W, 
Bagla:2009jy,% Seo:2009fq, 
2012A&A...540A.129A} and how well it can perform cosmological measurements \citep[see e.g.][]{Camera:2013kpa, Bull:2014rha}. The 21cm signal is also being used to study the Epoch of Reionization at higher redshifts, with several experiments running or in development (see e.g. \citealt{2013ExA....36..235M} for a review). As this line will be observed at radio frequencies, telescopes will naturally have lower resolutions, making this line an obvious application for intensity mapping. These telescopes also usually have large fields of view which allows them to cover large areas of the sky more quickly. Moreover, given the low frequencies, the HI line has very little background and foreground contamination from other lines, which is an advantage when comparing to intensity mapping with other lines as we will discuss later. Still, all lines have significant Galactic foregrounds that need to be removed or avoided. Several experiments have been proposed so far \citep{chime, 2013MNRAS.434.1239B, 2012IJMPS..12..256C, Bigot-Sazy:2015tot,Newburgh:2016mwi}, and some observations have already been made \citep{2011AN....332..637K}, including a detection in cross-correlation \citep{Chang:2010jp}. Moreover, a large HI intensity mapping survey has been proposed for SKA1-MID \citep{Santos:2015bsa}.

Although intensity mapping with the HI line holds great promises for cosmology, it is still relevant to ask if we can use other lines for such purposes. An obvious reason is that the 21cm line is quite weak. Although galaxy surveys with other lines have become routine, with detections made up to very high redshifts, the highest redshift detection for HI is $z=0.376$ \citep{2016ApJ...824L...1F}. Some studies have already been done for other lines, in particular for the Epoch of Reionization, such as  Ly$\a$ \citep{Silva:2012mtb}, CII \citep{Gong:2011mf}, CO \citep{Gong:2011ts}, H$_2$ \citep{Gong:2012iz} and others \citep{Visbal:2010rz}. At lower redshifts, \citet{Pullen:2013dir} have considered Ly$\a$ to probe the underlying dark matter power spectrum in a fiducial experiment, although they assume a very strong signal coming from the intergalactic medium (IGM). In fact, $\lya$ IM has already been detected in cross-correlation with quasars \citep{Croft:2015nna}. \citet{Uzgil:2014pga} considered the 3D power spectrum of CII and other far-infrared (FIR) emission lines. Similarly, \citet{Pullen:2012su} used the large scale structure distribution of matter to study CO emission while \citet{Breysse:2014uia} studied the possibility of using the first rotational transition of CO to do a survey at $z\sim3$. 

This paper presents a systematic study of all lines (besides HI) that can in principle be used for intensity mapping with reasonable experimental setups. It also compares them to determine which are the optimal lines to target for intensity mapping. It uses the latest observational and simulation results to predict the strength of the intensity and the expected bias towards dark matter. It then discusses the feasibility of surveys with these lines and which ones are more appropriate for cosmological applications. This is particularly relevant as the combination of different lines can bring exquisite constraints on large scale effects, such as primordial non-Gaussinanity, using the multi-tracer technique \citep{Seljak:2008xr, 2015PhRvD..92f3525A, 2015ApJ...812L..22F}. Moreover, the cross-correlation can be particularly useful in dealing with foregrounds/backgrounds as well as systematics. Finally, even if such intensity mapping surveys are not extremely competitive for cosmology, the simple detection of such a signal and its power spectrum will bring invaluable information about the astrophysical processes involved in the production of such lines and the clustering of the corresponding galaxies. 

We start with a review of the basic calculations for the average observed intensity of an emission line in Section \ref{sec:reviewIM}. In Section \ref{sec:lineemission} we model line emission in terms of the star formation rate of a given halo and estimate the signal for different lines. We then compare lines in Section \ref{sec:line_comp} so that we have an indication of which surveys to target in Section \ref{sec:Surveys}. The two following sections are focused on briefly assessing other sources of emission/contamination and address how to deal with such issues. We conclude in section \ref{sec:discussion}.

\section{Line Intensity mapping - Review} \label{sec:reviewIM}

\subsection{Coordinates and Volume factors}
Let us consider a volume in space with comoving centre {\boldmath $\chi_0$}, two angular directions perpendicular to the line of sight, and another direction along the line of sight. Using the flat sky approximation the comoving coordinates of a point inside such volume will be
\be
{\boldsymbol r} = {\boldsymbol \chi_0} + \D\theta_1 D_A\hat{\boldsymbol \theta}_1 + \D\theta_2 D_A \hat{\boldsymbol \theta}_2 + \D\nu \tilde y \hat{\boldsymbol r}\,.
\ee \unboldmath
The first two components are the displacements perpendicular to the line-of-sight, where $D_A$ is the angular diameter distance in comoving units (for a flat universe is just the comoving distance $\chi$). The last component corresponds to the line-of-sight or the redshift/wavelength/frequency direction. A small variation in the comoving distance corresponds to a variation in the observed frequency as $\D\chi=\tilde y~\D\nu$, where $\tilde y$ is defined by
\be \label{ydef}
\tilde y\equiv \frac{d\chi}{d\nu}=\frac{\la_e\l 1+z\r^2}{H\l z\r}\,.
\ee  
Here $\la_e$ is the wavelength of the emitted photon and $H$ is the Hubble parameter at the redshift $z$, so that $\la_O=\la_e(1+z)$.

\subsection{Average observed signal}

Let us assume that galactic emission lines have a sharp line profile $\psi(\la)$ that 
can be approximated by a Dirac delta function around a fixed wavelength $\la$, i.e., $\psi(\la)\simeq \delta^D\l \la_e-\la\r$. This way we can assume a one-to-one 
relationship between the observed wavelength and redshift. This approximation is valid as long as the observational pixel is considerably larger than the FWHM of the targeted line. % and the effect of peculiar velocities. 
Although this will generally be the case in intensity mapping experiments, we will discuss the implications of a non-sharp line profile in Section \ref{sec:broad_lines}. 

The average observed intensity of a given emission line is given by
\be \label{eq:savehm}
\bar I_\nu (z)=\int_{M_{min}}^{M_{max}} \d {\rm M}~\hmf ~\frac{L(M,z)}{4\uppi D_L^2}~\tilde y D_{A}^2 \,,
\ee
where $dn/dM$ (Halos/Mpc$^3$/$M_\odot$) is the comoving halo-mass function (HMF) \citep{Sheth:1999mn}, $D_L$ is the 
luminosity distance ($D_L=\l 1+z\r \chi$ for a flat universe) and $\tilde y D_{A}^2$ is the comoving unit volume of the 
voxel. Note that this is the physical average intensity and with the definition of $\tilde y$ (Eq. (\ref{ydef})) its 
units are power per unit area per unit solid angle per unit frequency. The galaxy line luminosity $L\l M,z\r$ 
depends on the halo mass and often evolves with redshift. The relevant mass interval of 
integration depends on the chosen line and is redshift dependent. We assume $10^8M_{\odot}$ as the minimum mass 
of a halo capable of having stellar formation due to atomic cooling. If we instead assumed stellar formation 
through molecular cooling, then the minimum mass would be considerably lower \citep[down to $\sim 10^6 M_{\odot}$;][]{Visbal:2014fta}. 
If not otherwise stated, we will consider the dark matter halos mass in the range $[10^8,10^{15}] ~\rm M_\odot$. 

Alternatively, one can estimate the average intensity of a line using luminosity functions via
\be \label{eq:ILF}
\bar I_\nu \l z\r=\int_{L_{\rm min}/L^*}^{L_{\rm max}/L^*} \d\l\frac{L}{L^*}\r ~\frac{L}{4\uppi D_L^2}~\phi\l L\r ~\tilde y D_{A}^2\,.
\ee  
We take the luminosity function (LF) $\phi$ to have the Schechter form
\be \label{eq:LF}
\phi\l L\r d\l\frac{L}{L^*} \r= \phi^* \l \frac{L}{L^*}\r^\a e^{-L/L^*}d\l\frac{L}{L^*}\r\,
\ee
where $\alpha$ is the slope of the faint end, $L^*$ is the turn over luminosity for the bright end decay and $\phi^*$ is the overall normalization of the luminosity function. The set of parameters $\{\phi^*,L^*,\a\}$ depend on the line in consideration as well as the integration limits.

\subsection{Expected signal from broad emission lines}
\label{sec:broad_lines}
In some cases the considered emission line has a broad line profile due to the motion of the gas in the galaxy. This causes 
the previous discussion to breakdown as the signal from a galaxy will be observed at several pixels along a line-of-sight. 
To minimize this issue one could consider frequency bins which are at least of the same order of magnitude as the FWHM of the line. Alternatively, one could estimate the amount of signal observed at each pixel using the line profile. In this case, the galaxy will be responsible for the flux 
observed in several pixels. Let us assume we have a galaxy at redshift $z_c$ and that the line in study has total luminosity $L$ and a line profile $\psi (\lambda_e)$. 
Since $\int \psi (\lambda_e) d\lambda_e=1$ we can write the luminosity dependence as 
\be
L_{line}(\lambda_e)=L ~ \psi(\lambda_e)\,.
\ee
Due to the line profile, galaxies at different redshifts will contribute to the observed signal at a given wavelength. Therefore, the 
intensity at an observed wavelength will be an integral over all redshifts weighted by the line profile. Hence, the expected observed flux will be
\bea
\label{Sbline}
\bar I_\nu (\la_O)=\int {\rm dz} \frac{\la_O}{\l 1+z\r^2} \int_{M_{min}}^{M_{max}} dM~\hmf\times\qquad&&\nn\\
\times ~ \frac{L(M,z) \psi(\la_O/(1+z))}{4\uppi D_L^2 \l z\r} ~ \tilde y \l z\r D_{A}^2\l z\r \,. &&
\eea

Such broadening of the emission line washes away small scale fluctuations but one would not expect large effects on large angular 
scales.
For the purpose of this paper we will consider redshift bins larger than the broadening of the 
line, effectively approximating the emission line profile by a delta function. 

\section{Atomic and molecular emission lines from galaxies} \label{sec:lineemission}

\subsection{Relation between line luminosity and SFR}

Our aim is to study the power spectrum of matter perturbations up to redshift $z=5$ by performing 
intensity mapping of the strongest emission lines from galaxies. Most of the lines we consider are sourced by ultra-violet (UV) emission from young stars and quasars. A fraction of the UV emission is absorbed by galactic dust which in turn will produce a strong continuum thermal emission which contributes to the infrared continuum background. Another fraction of the ionizing UV radiation produced by galaxies escapes into the IGM, while a further part of this radiation is absorbed by the galaxy gas, heating it and exciting and/or ionizing 
its atoms and molecules. As the gas recombines and/or de-excites, transition lines are emitted from the gas.  In most cases the recombination time is significantly shorter 
than a hubble time so we can assume that the process is immediate. As a result the emission rate of a recombining line will be proportional to the stellar ionizing flux, the source of which is the star formation in the galaxy. It therefore relates to the star formation rate (SFR) i.e., the amount of mass transformed into stars per year. This dependence is not necessarily linear and needs to be calibrated 
for each line. We further need to consider that only a portion of the line luminosity exits the galaxy. 
We then model the total luminosity associated with a halo of mass $M$ as
\be \label{eq:LSFR}
L =K\l z\r \times  \l \frac{SFR \l M,z\r}{\rm M_\odot /yr}\r^\gamma \,,
\ee
where $K(z)$ includes the calibration of the lines as well as the several escape fractions, while the parameter $\gamma$ is 
introduced for those lines that do not scale linearly with the SFR. This generic model was chosen to accommodate all the details of the lines that we will study in this paper.

The star formation rate density (SFRD) is defined as 
\be
\label{eq:sfrd}
SFRD\l z\r\equiv\int{\rm dM}~SFR\l M,z\r ~\hmf\,,
\ee
where the integration is done over all dark matter halos and has units of ${\rm M_{\odot}\, yr^{-1}\, Mpc^{-3}}$. In this study we attempt to cover the uncertainty in the SFRD by using two different SFR models.

\begin{table*}
\centering            
\caption{Fit to the SFR parameters of Eq. \ref {eq:sfrsmill} based in the average relations from simulated galaxy catalogs obtained by \citet{DeLucia:2006vua} and \citet{Guo:2010ap} who post processed the Millennium I \citep{Springel:2005nw} and II \citep{BoylanKolchin:2009nc} simulations respectively.}            
\begin{tabular}{l  c c c c c c}        
\hline\hline                 
Redshift  & $M_{\rm 0}$ & $M_{\rm b}$ & $M_{\rm c}$ & $a$ & $b$ & $c$\\    
\hline                 
   0.0 & $3.0\times10^{-10}$  & $6.0\times10^{10}$   & $1.0\times10^{12}$  & 3.15  & -1.7 & -1.7\\
   1.0 & $1.7\times10^{-9}$   & $9.0\times10^{10}$   & $2.0\times10^{12}$  & 2.9  & -1.4 & -2.1\\
   2.0 & $4.0\times10^{-9}$   & $7.0\times10^{10}$   & $2.0\times10^{12}$  & 3.1  & -2.0 & -1.5\\
   3.0 & $1.1\times10^{-8}$   & $5.0\times10^{10}$   & $3.0\times10^{12}$  & 3.1  & -2.1 & -1.5\\
   4.0 & $6.6\times10^{-8}$   & $5.0\times10^{10}$   & $2.0\times10^{12}$  & 2.9  & -2.0 & -1.0\\
   5.0 & $7.0\times10^{-7}$   & $6.0\times10^{10}$   & $2.0\times10^{12}$  & 2.5  & -1.6 & -1.0\\
  \hline                                  
\end{tabular}
\label{tab:SFR_param}     
\end{table*}

The first model we consider uses the Markov Chain Monte Carlo (MCMC) method to estimate the galaxies' stellar mass versus galaxy mass relation as a function of redshift based on several observational constraints in the literature \citep{Behroozi:2012iw}. We will refer to this model as Be13. They provide the SFR as a function of halo mass and redshift, which can be found and downloaded from the authors' personal page\footnote{http://www.peterbehroozi.com/data.html}. They also studied the literature to produce a new best fit the the SFRD which reads
\be \label{eq:sfrdbe12}
SFRD\z=\frac{0.180}{10^{-0.997\l z-1.243\r}+10^{0.241\l z-1.243\r}}\,.
\ee
 Please note that when we computed the SFRD with Be13 SFR(M,z) and the HMF using Eq. (\ref{eq:sfrd}) we found a discrepancy between Be13 best fit to the SFRD and our estimate to the sFRD using Be13 SFR(M,z). Hence we corrected the SFR by a redshift dependent fraction such that we recover Eq. (\ref{eq:sfrdbe12}) when using our HMF in Eq. (\ref{eq:sfrd}). We use a second model to account for the large uncertainty in SFR observations at high redshifts ($z>2$). This model is a parameterization of simulated galaxy 
catalogs, which we will refer to as SMill. The galaxy catalogs used were obtained by \citet{DeLucia:2006vua} and 
\citet{Guo:2010ap} who post processed the Millennium I \citep{Springel:2005nw} and II \citep{BoylanKolchin:2009nc} simulations respectively. The SFR parameterization in terms of 
the mass and redshift is given by
\be\label{eq:sfrsmill}
SFR (M,z)= M_0 \l \frac{M}{M_a} \r ^{a} \l 1+ \frac{M}{M_b}\r^{b} \l 1+\frac{M}{M_c}\r^{c}\,,
\ee
where $M_a=10^8 {\rm M}_{\odot}$ and, $M_0$, $M_b$, $M_c$, $a$, $b$ and $c$ evolution with redshift is given in table \ref{tab:SFR_param}.

\begin{figure}
\begin{centering}  
\includegraphics[angle=0,width=\columnwidth]{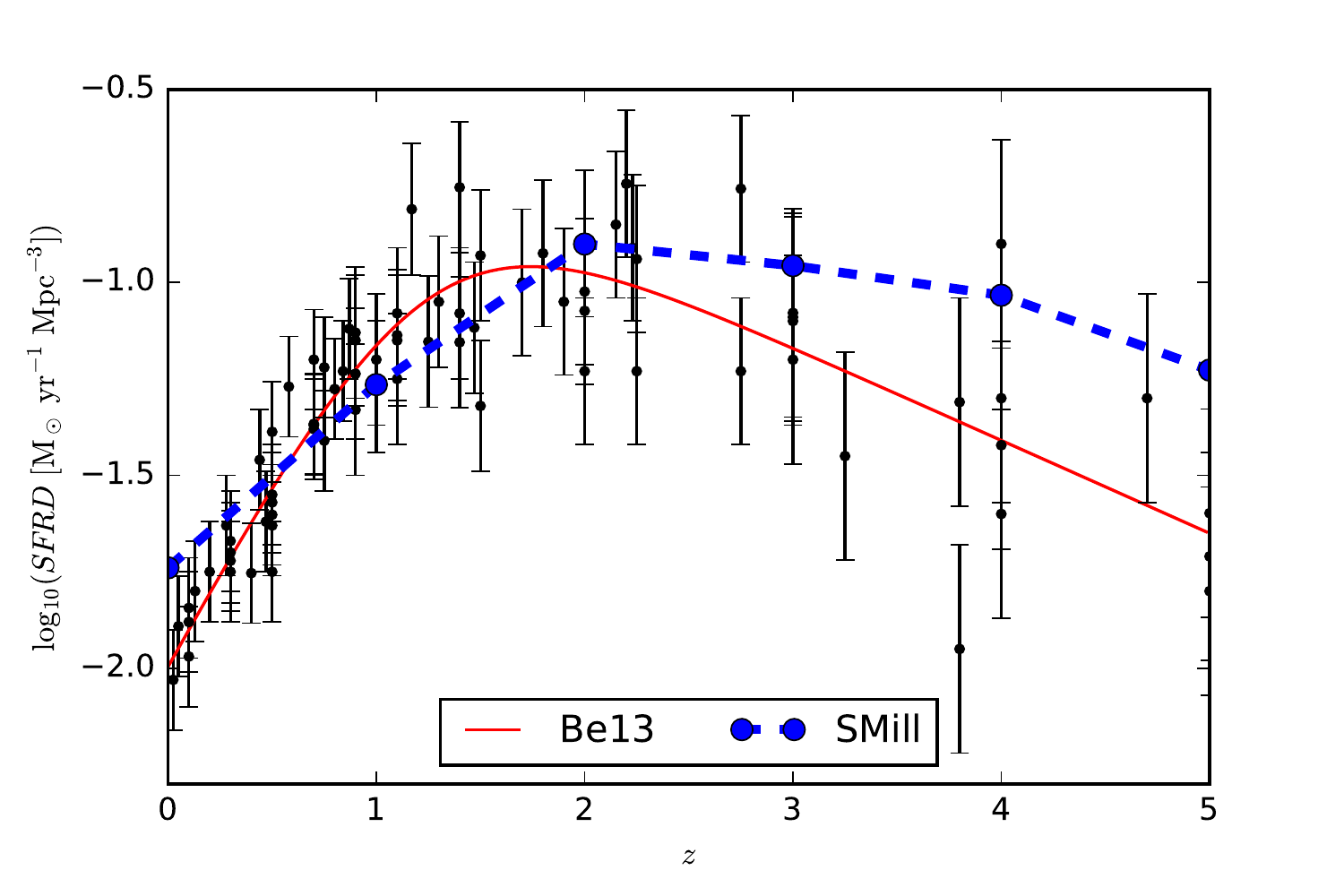}
\caption{SFRD redshift evolution for Be13 model \citep{Behroozi:2012iw} in solid red and for SMill model \citep{DeLucia:2006vua,Guo:2010ap} in thick-dashed blue. The black dots are a recollection of observational estimates of the SFRD in the literature systematized by \citet{Behroozi:2012iw}.}
\label{fig:sfrd}
\end{centering}
\end{figure}

In figure \ref{fig:sfrd} we compare the time evolution of the SFRD predicted by the two models described above. We also 
plot as black bold dots the data points from \citet{Behroozi:2012iw}. The split in the two
models at $z=2$ reflects the uncertainty in this quantity due to the lack of reliable observations at higher redshifts. We thus consider Be13 as a lower estimate of the SFR while SMill will be taken as a upper estimate.

It is conventional to study the observed intensity with the quantity $\nu I_\nu$ which we will follow in this paper. 
For $\gamma=1$ one can use Eqs. (\ref{eq:savehm}), (\ref{eq:LSFR}) and assuming that all dark matter halos contribute to the signal, i.e., Eq. (\ref{eq:sfrd}) to get
\be \label{eq:nuI}
\nu \bar I_\nu \l z\r=  \frac {c~K\z ~ SFRD \l z\r}{4\uppi \l1+z\r H\l z\r} \,.
\ee
We will use this quantity such that the units of $\nu \bar I_\nu$ are given in erg/s/cm$^2$/Sr.
This quantity can also be determined using luminosity functions. Defining the luminosity density as
\be \label{eq:rhoL}
\bar \rho_L \l z\r\equiv\int_{L_{min}/L^*}^{L_{max}/L^*} d\l\frac{L}{L^*}\r ~L~\phi\l L\r\,,
\ee
it follows that
\be \label{eq:nuILF}
\nu \bar I_\nu \l z\r=\frac{c ~\bar\rho_{L}\l z\r}{4\uppi \l 1+z\r H\l z\r}  \,.
\ee

\subsection{Lyman-$\a$}

The hydrogen Lyman-$\a$ emission line is the most energetic line coming from star forming galaxies. $\lya$ is a 
UV line with a rest wavelength of 121.6 nm. It is mainly emitted during hydrogen recombinations, although it can 
also be emitted due to collisional excitations. As the recombination timescale for hydrogen is usually small, we 
assume that the number of recombinations is the same as the number of ionizations made by UV 
photons emitted by young stars in star forming galaxies. These photons have neither been absorbed by the galactic dust nor escaped the 
galaxy. The emission of $\lya$ photons is in fact reprocessed UV emission that ionized the neutral hydrogen of the galaxy. 
As the Lyman-$\alpha$ photons are emitted and travel through the interstellar medium (ISM) they get absorbed and re-emitted by neutral 
hydrogen until they escape the galaxy. This scattering causes the photons' direction to change randomly, hence 
there is a negligible probability that photons retain their initial direction. Such a random walk in the ISM 
of the galaxy also increases the probability that the $\lya$ photons are absorbed by dust. The galaxy 
metallicity and its dust content therefore has an impact in the observed Lyman-$\alpha$ emission. Similarly, peculiar motion 
of the hydrogen gas broadens the line width. Thus, in an intensity mapping survey the redshift resolution must be higher than the FWHM of the emission line at a particular redshift.

It is common in the literature to assume that $\lya$ emission is linear in the SFR \citep[e.g.:][]{Kennicutt:1998zb,Ciardullo:2011vj}, i.e., $\gamma_\lya=1$. Therefore, one can estimate the $\lya$ signal using Eq. (\ref{eq:nuI}). We model $K^\lya$ in Eq. (\ref{eq:LSFR}) as:
 \be \label{eq:Kz}
K^{\lya}\l z\r =\l f_{dust}^{\rm UV}-f_{esc}^{\rm UV}\r \times f^{\lya}_{esc}\l z\r\times R^{\lya}  \,,
\ee
where $f_{dust}^{\rm UV}$ is the fraction of UV photons that are not absorbed by dust, $f_{esc}^{\uv}$ is fraction of UV photons that escape the star forming galaxy, $f^{\lya}_{esc}$ is the fraction of $\lya$ photons that escape the galaxy and $R^{\lya}$ is a constant in units of luminosity which calibrates the line emission.

The probability of a Lyman alpha photon being emitted during a recombination is high, and assuming an optically thick interstellar medium, Case B 
recombination, and a \citet{1955ApJ...121..161S} universal initial mass function we have \citep{Ciardullo:2011vj}
\be \label{Rlya}
R_{rec}^\lya = 1.1 \times 10^{42} ~ {\rm erg/s}\,.
\ee
Although this is the conventional relation to connect a galaxy SFR with its $\lya$ luminosity, in the low redshift universe 
the use of a Case A recombination coefficient might be more appropriate if the neutral gas is restricted to very dense 
regions. In that case the recombination rate would be higher but the Lyman-$\a$ luminosity would be lower. Case A refers to the case when recombinations take place in a medium that is optically thin at all photon frequencies. 
On the other hand case B recombinations occur in a UV opaque medium, i.e, optically thick to Lyman series and ionizing photons, where direct recombinations to the ground state are not aloud.
%On the other hand, when recombinations occur in an UV opaque medium, i.e., optically thick to Lyman series and ionizing photons, then we call it Case B. 
For a review please refer to \citet{Dijkstra:2014xta}. Collisional excitations give an extra contribution which is an order of magnitude lower, $R_{exc}^{\lya} = 4.0 \times 10^{41}$ erg/s, at a gas temperature of $10^4K$. For now we will ignore this contribution since it was not considered in the estimation of an observationally determined $f_{esc}^{\lya}$. 

\subsubsection{UV dust absorption and escape fraction}
In Eq. (\ref{eq:Kz}) the terms in brackets correct the intrinsic luminosity for the fact that only a fraction of the 
UV photons produced by young stars are consumed in the ionization of ISM gas clouds. I.e., a fraction of these photons 
escape the galaxy or are absorbed by dust. The values of $f_{esc}^{\uv}$ and $f_{dust}^{\uv}$ for a single galaxy depend 
on physics at 
both large (SFRs and gas masses, etc) and very small scales (gas clumping). Accurate modeling of these quantities is thus 
out of the reach of simulations. However, observational measurements of $f_{esc}^{\rm UV}$ have been done 
for a few galaxies and only along a few lines of sight \citep{Shapley:2006fp, Siana:2010gp, Nestor:2011iq}. Also, its estimation is dependent on measurements of the astrophysical conditions in the inter-galactic medium (IGM), which on its own depends on the intensity of the UV radiation which successfully escaped the 
galaxy. It is 
therefore an indirect probe of $f_{esc}^{\uv}$ which has so far provided contradictory values. 
Hence, in this paper we will consider $f_{esc}^{\uv}$ as a cosmological average of the percentage of UV photons that escape a galaxy and $f_{dust}^{\uv}$ as a cosmological average of the percentage of UV photons that are not absorbed by dust in the galaxies. In this communication we will assume that $f_{esc}^{\uv}=0.2$ \citep{2014MNRAS.440..776Y}. We will follow the conventional assumption that dust gives an average extinction of 1 mag, i.e., 
\be \label{eq:fesc_ext}
f_{\rm dust}^\uv=10^{-E_\uv/2.5}\,.
\ee       
Following \citet{Sobral:2012fb}, we will consider that $0.8<E_{\rm UV}<1.2$ for our lower and higher estimates of the signal. As a first approximation we will consider both quantities to be constant with redshift. We note that these factors fluctuate according to the galaxy properties, however for intensity mapping studies 
we are averaging over several galaxies and so the intrinsic dispersion will not be a problem.

\subsubsection{$\lya$ escape fraction}
The cosmological average of the percentage of $\lya$ photons that escape the galaxies is also very challenging to determine 
observationally or through simulations. The observationally based model for the escape fraction of Lyman alpha photons 
from \citet{Hayes:2010cj} is given by 
\be
f_{esc}^{\lya}= C \l 1+z\r^{\xi}\, 
\ee
with $C=\l 1.67^{+0.53}_{-0.24}\r \times 10^{-3}$ and $\xi=2.57^{+0.19}_{-0.12}$. The parameter values were estimated using 
observational data of the $\lya$ and $\ha$ relative intensities at low redshift, as well as the $\lya$ intensity relative to 
the continuum UV emission at high redshifts. The low values of $f_{esc}^{\lya}$ at low redshift are likely to be due to the 
$\lya$ and $\ha$ ratio being inversely proportional to galaxy luminosity. We also note that the calibration at low redshifts 
was made accounting for dust absorption and possible absorption/scatter in the IGM regardless of the production mechanism of 
$\lya$ photons. At high redshifts the calibration was made comparing the expected $\lya$ emission from recombinations with 
the observed $\lya$ emission, which included collisional excitations as a source of $\lya$. All of these effects result in a 
lower estimate of $f_{esc}^{\lya}$. Other authors have studied the Lyman-$\a$ escape fraction 
\citep{Dijkstra:2013zxa,Ciardullo:2014cba} arriving to similar results using luminosity functions. These 
are especially dependent on the LF low-end cut-off and fainter galaxies might have higher escape fractions. These estimations 
not only determine the escape fraction due to extinction, but also from scattering of photons in the galaxy. This 
is of particular importance for IM, since all emission is integrated out in a single pixel including the scattered photons 
that end up leaving the galaxy anyway. A significant fraction of the overall $\lya$ emission is likely to be produced by 
sources too faint to be observed with current technology, therefore raising the real value of $f_{esc}^{\lya}$. Another 
reason to read these results as underestimations of the escape fraction relies on the fact that such estimates are based 
on observations of Lyman-$\a$ emitters (LAEs). These are mainly sensitive to the bulge of the galaxy where the dust component 
is higher, and therefore have overall lower $\lya$ escape fractions. More recently \citep{Wardlow:2013qpa} estimated a substantially 
higher $f_{esc}^{Ly\a}$ at redshifts $z=2.8$, $z=3.1$ and $z=4.5$. Intensity mapping experiments have enough sensitivity to 
capture the $\lya$ photons scattered around the galaxy, so the Lyman-$\a$ escape fraction considered should only be due to 
dust absorption. Hence, it should be similar to the escape fraction of H$\a$ (see \ref{sec:ha}). This is around 40\% for the 
massive star forming galaxies. For the less massive systems, which are out of the reach of the sensitivity of most 
observational experiments, it may be even higher \citep{Dawson:2011kf,Price:2013mma}. In light of these considerations we will 
take the calibration of \citet{Hayes:2010cj} as a lower estimate for the escape fraction and $f_{esc}^{\lya}|_{max}=0.4$ 
constant in redshift as an upper estimate. At each redshift we will take $f_{esc}^{\lya}=0.2$ as our average value.

\subsubsection{$\lya$ Intensity}

\begin{figure}
\begin{centering}  
\includegraphics[angle=0,width=\columnwidth]{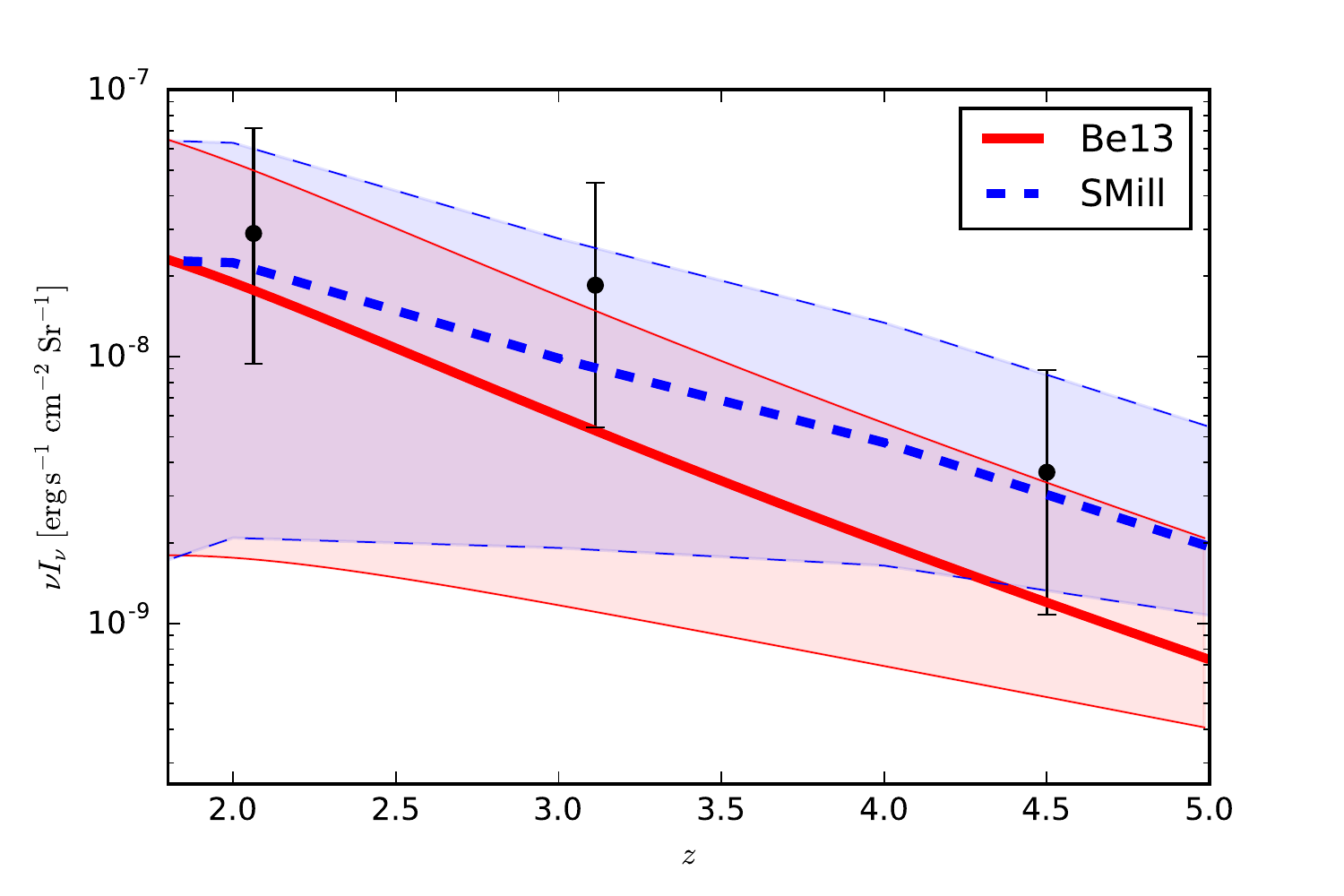}
\caption{Average $\lya$ intensity as a function of redshift. The thick solid red line correspond to estimates using the Be13 SFR model \citep{Behroozi:2012iw} and the thick dashed blue to the estimates obtained with SMill SFR model \citep{DeLucia:2006vua,Guo:2010ap}. The shaded regions encompass the uncertainties in $f_{dust}^\uv$ and $f_{esc}^\lya$ and are bounded by thinner lines. The black dots were computed using the LAE luminosity functions given by \citet{Guaita:2009nj} at $z=2.063$, \citet{Ciardullo:2011vj} at $z=3.113$ and \citet{Zheng:2011iz} at $z=4.5$.}
\label{fig:nuI:lya}
\end{centering}
\end{figure}

In Figure \ref{fig:nuI:lya} we compare the estimates of the average Ly$\a$ intensity. The lines show estimates of the expected intensity for the SFR models Be13 in red and SMill in blue computed using Eq. (\ref{eq:nuI}). We also considered variation due to the escape fraction of Lyman-$\a$ as well as the UV escape fraction. The dots are estimations of the average intensity computed using Eq. (\ref{eq:ILF}) and the luminosity functions (Eq. \ref{eq:LF}) of LAEs calibrated by \citet{Guaita:2009nj} at $z=2.063$, \citet{Ciardullo:2011vj} at $z=3.113$ and \citet{Zheng:2011iz} at $z=4.5$. One should note that no further correction needs to be introduced in this estimation since LAEs LF are already observational. One can see that they agree within the uncertainties considered. Nonetheless, the LF estimates are systematically near the higher bounds of our estimates using the SFR. One of the reasons for this comes from the fact that $\lya$ escape fractions are calibrated using the emission from the bulge of galaxies which has much more extinction than the edges of the galaxy.
\

Ly$\a$ emission which arrives in the optical will be contaminated by lower redshift foregrounds as OII [373.7 nm], the OIII doublet [495.9 nm and 500.7 nm] and 
other fainter metal lines as well as Balmer series lines. We will discuss the most relevant of these lines in the following subsections. Another contaminant to be taken into account is the UV continuum background emission which will be originated by young stars and quasars at higher redshifts. Although there is around one order of magnitude uncertainty, its intensity will definitely be higher than Ly$\a$ emission \citep{Dominguez:2010bv}. We will not address this problem in this paper but we note that its power spectrum should be nearly flat.

\subsection{$\ha$} \label{sec:ha}

The UV light emitted by young stars ionizes the surrounding gas in the ISM. As hydrogen recombines in a cascading process, several lines other than $\lya$ are emitted, such as H$\alpha$. This is the lowest line of the Balmer series, with a rest-frame wavelength of 656.281 nm. Just like $\lya$, we will assume that the luminosity of $\ha$ is linear in the SFR ($\gamma_\ha=1$), and that $K^\ha\l z\r$ follows Eq. (\ref{eq:Kz}). We consider the same values and uncertainties for $f_{esc}^{\uv}$ and $f_{dust}^{\uv}$. \citet{Kennicutt:1998zb} assumed an optically thick interstellar medium, Case B recombination, a Salpeter universal initial mass function and that all UV continuum is absorbed by the gas in the galaxy, arriving at
\be \label{Rha}
R^\ha = 1.3 \times 10^{41} {\rm erg/s} \,.
\ee      
To estimate the observed H$\a$ luminosity we also need to account for dust absorption. \citet{Kennicutt:1998zb} assumed an extinction $E_{\rm H\a}\sim 1$ mag. Several authors have studied H$\a$ extinction in galaxies and produced similar results \citep{James:2004ha,Sobral:2012fb}. Most H$\a$ studies cannot probe the redshift evolution of $E_{\rm H\a}$.  Here we take a conservative approach and assume the extinction to be roughly constant. One should bear in mind that at higher redshifts, we expect a higher signal than predicted here, given the lower dust content. As for UV light, the H$\a$ escape fraction will be given by Eq. (\ref{eq:fesc_ext}) with $E_\ha=1$ mag (roughly 40$\%$) with an ``uncertainty" of $0.2$ magnitudes.

\begin{figure}
\begin{centering}  
\includegraphics[angle=0,width=\columnwidth]{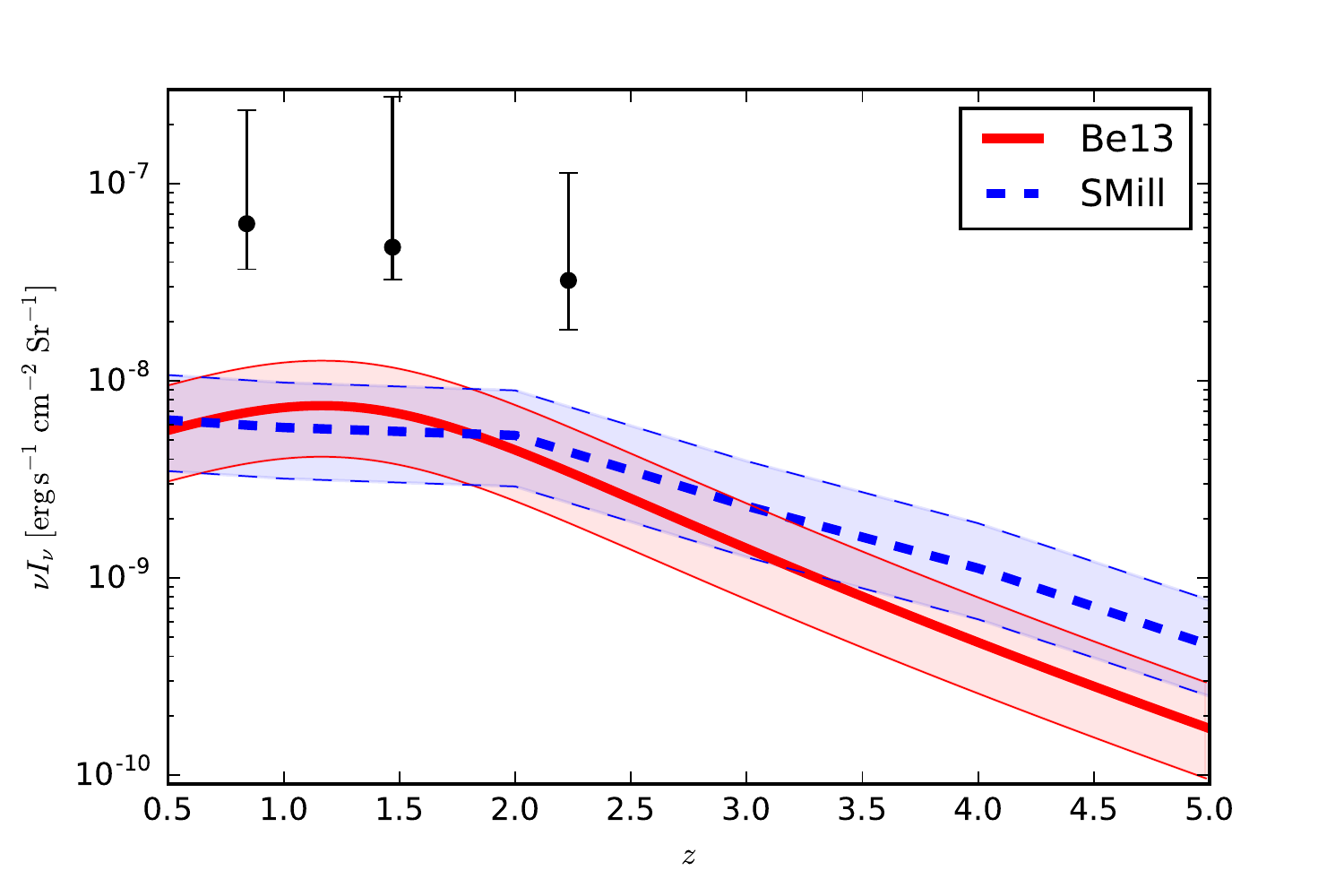}
\caption{Average $\ha$ intensity as a function of redshift. 
The thick solid red line correspond to estimates using the Be13 SFR model \citep{Behroozi:2012iw} and the thick dashed blue to the estimates obtained with SMill SFR model \citep{DeLucia:2006vua,Guo:2010ap}. The shaded region encompass the uncertainties in $f_{esc}^\uv$ and $E_\ha$ and are bounded by thinner lines. The black dots were computed using the $\ha$ luminosity functions given by  \citet{Sobral:2012fb} at $z=0.8$, $z=1.47$ and $z=2.23$, and corrected with $f_{esc}^\ha$.}
\label{fig:nuI:ha}
\end{centering}
\end{figure}

We compare our estimates for the average intensity of $\ha$ as a function of redshift in figure \ref{fig:nuI:ha}. The thick solid red line shows our estimates computed using Be13 model, while the thick blue dashed lines uses SMill model in Eq. (\ref{eq:nuI}). The dots are estimates of the average intensity computed using Eq. (\ref{eq:ILF}) and $\ha$ luminosity functions determined by \citet{Sobral:2012fb}, and corrected for extinction. One should note these are intrinsic $\ha$ luminosity functions rather than observed luminosities,as was the case for LAEs. Although one can argue that the estimates still lie within the same order of magnitude, we notice that the LF estimates are systematically higher. We are either underestimating the SFR from galaxies or the LF are over estimating the signal. Another possibility is related to the value of extinction that $\ha$ photons experience. We used calibrations that are generically obtained by looking at the bulge of galaxies. These have higher extinctions than the ones experienced at the edges of the galaxy. Another possible explanation comes from the amount of UV that sources $\ha$ emission which we may be under-estimating. We will not explore such discrepancies further since our goal is to study the feasibility of IM with galactic emission lines.

For completeness, one should indicate which are the strongest contaminants of $\ha$. Background contamination comes mainly from oxygen lines (the OII [373.7 nm] and the OIII doublet 
[495.9 nm and 500.7 nm]), NII[655.0 nm] and other lines from the Balmer series, but is mainly due to UV lines in the Lyman series. Foregrounds come mainly from continuous infra-red (IR) emission from galaxies.

\subsection{H$\beta$}

H$\beta$ is the second strongest line of the Balmer series, with a rest emission wavelength of 486.1 nm. In the same way as $\ha$, H$\beta$ follows Eq. (\ref{eq:LSFR}) with $\gamma_{\rm H\beta}=1$ and 
\be \label{Rhb}
R^{\rm H\beta} = 4.45 \times 10^{40} ~{\rm erg/s} \,.
\ee
This is a well known result since $[H\beta/\ha]=0.35$ for optically thick interstellar medium, Case B recombination and a Salpeter universal initial mass function. We use the same UV escape fractions as before, but assume that the H$\beta$ extinction is slightly higher than for $\ha$, i.e., $E_{\rm H\beta}=1.38$ mag \citep{Khostovan:2015aa}. The signal estimations for H$\beta$ are similar to the ones shown in figure \ref{fig:nuI:ha} except they are rescaled by an $[H\beta/\ha]f_{esc}^{\rm H\beta}/f_{esc}^{\ha}$ factor.

\subsection{Oxygen lines}

\subsubsection{Rest-frame emission in the optical}

\begin{figure}
\begin{centering}  
\includegraphics[angle=0,width=\columnwidth]{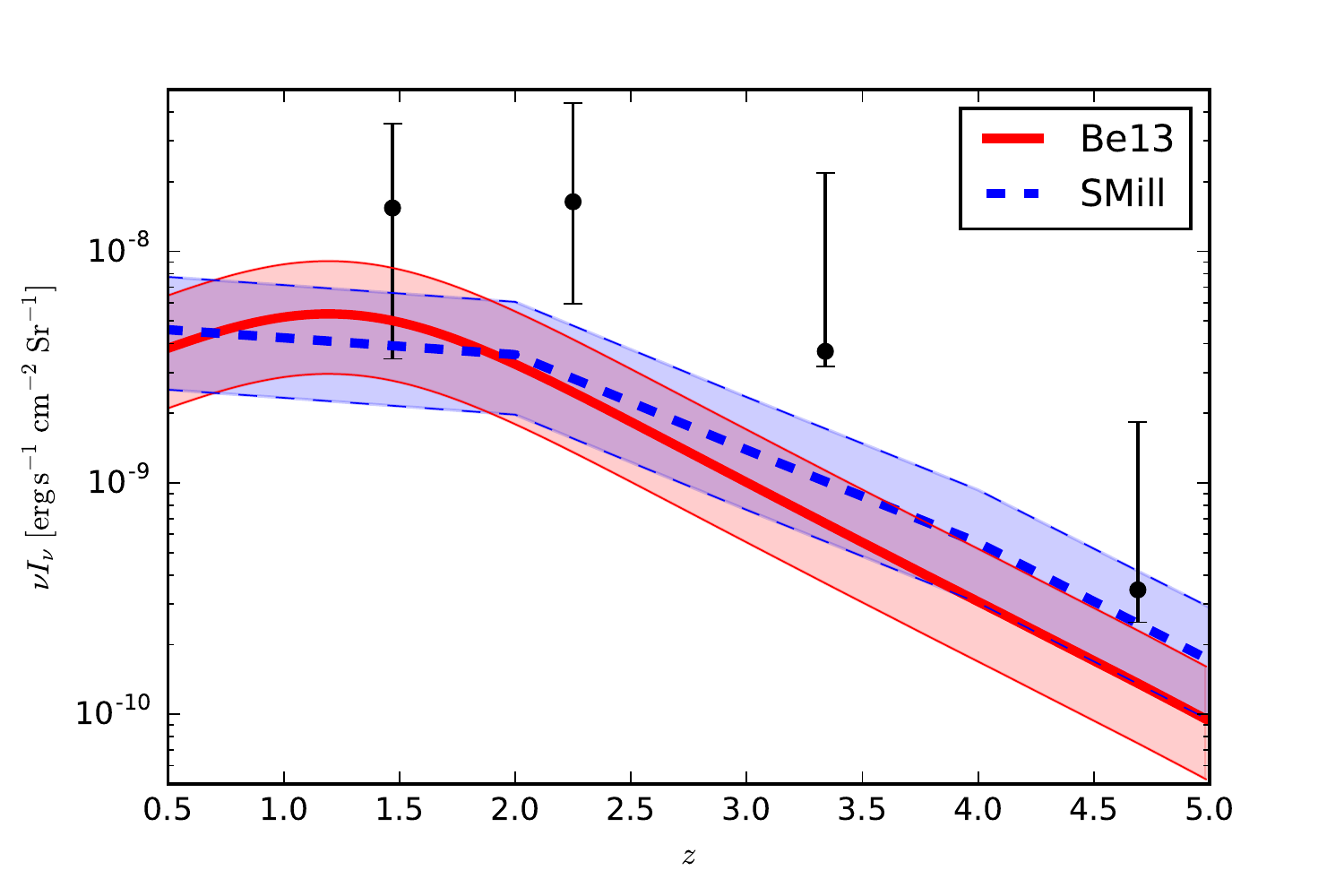}
\caption{Average OII intensity as a function of redshift. The thick solid red line correspond to estimates using the Be13 SFR model \citep{Behroozi:2012iw} and the thick dashed blue to the estimates obtained with SMill SFR model \citep{DeLucia:2006vua,Guo:2010ap}. The shaded region encompass the uncertainties in $f_{esc}^\uv$ and $E_{\rm OII}$. The black dots were computed using the OII luminosity functions given by \citet{Khostovan:2015aa} at $z=1.47,2.25,3.34,4.69$, and corrected with $f_{esc}^{\rm OII}$.}
\label{fig:nuI:oii}
\end{centering}
\end{figure}

Ionized oxygen produces several emission lines which contribute to a galaxy spectrum in the visible and FIR. The forbidden optical oxygen line OII[372.7 nm] is a strong emission line which has been used by the Sloan Sky Digital Survey [http://www.sdss.org] to study galaxies and determine their redshifts. As for previous UV and optical lines, we take OII luminosity to be linearly dependent in the SFR, i.e., $\gamma_{\rm OII}=1$. The line strength has been estimated by \citet{Kennicutt:1998zb}, assuming a Salpeter (1955) universal initial mass function and Case B recombinations, as
\be \label{Roii}
R^{\rm OII} = 7.1 \times 10^{40}  ~{\rm erg/s} \,.
\ee
The escape fraction of OII, $f_{esc}^{\rm OII}$, is given by Eq. (\ref{eq:fesc_ext}) with $E_{\rm OII}=0.62$ \citep{Khostovan:2015aa}, with an extinction uncertainty of 0.2 mag. The values of $f_{esc}^{\uv}$ and $f_{dust}^{\uv}$ are the same as the ones used in previous emission lines. To compute the intensity we used Eq. (\ref{eq:savehm}) with the mass integration range $[10^{11},10^{15}] ~\rm M_\odot$. We expect lower mass halos to be metal poor \citep{Henry:2013yfa}, hence the assumed mass cut-off. %lack of observations of low mass galaxies, we use a mass cutoff which corresponds to a luminosity limit one order of magnitude below the currently available observations. We do not include low mass halos since one expect them to have very low metallicities \cite{Henry:2013yfa}.
In figure \ref{fig:nuI:oii} we compare the different estimates of the OII signal. The thick lines show the estimates computed using Be13 (solid red) and SMill (dashed blue) models using the quoted values for $f_{esc}^\uv$, $f_{dust}^\uv$, $f_{esc}^{\rm OII}$ and $R^{\rm OII}$. The shaded region corresponds to variations in the escape fractions, while the black dots are estimates using \citet{Khostovan:2015aa} OII luminosity functions, and corrected for extinction. Our estimates are within the same order of magnitude as the ones using LF. This agreement is strongly dependent on the chosen minimum integration mass. Here we do not included the metallicity dependence on redshift and halo mass, since it is poorly known. This may explain the discrepancies, but our approximation still describes the trend of the signal. Such metallically dependence is beyond the scope of this paper but one should bear in mind that this problem needs to be addressed for a proper use of OII for IM. %Up to $z=5$ OII is a low redshift contaminant to Lyman-$\alpha$ and a background contaminant to H$\alpha$, except for a small redshift window in which we expect its signal to be dominant. Within the range where it is the dominant emission, optical ground telescopes could be redesigned to perform OII intensity mapping for $z<1$. This would just complement current surveys and strength the possible uses of the technique as well as improvements. 

\begin{figure}
\begin{centering}  
\includegraphics[angle=0,width=\columnwidth]{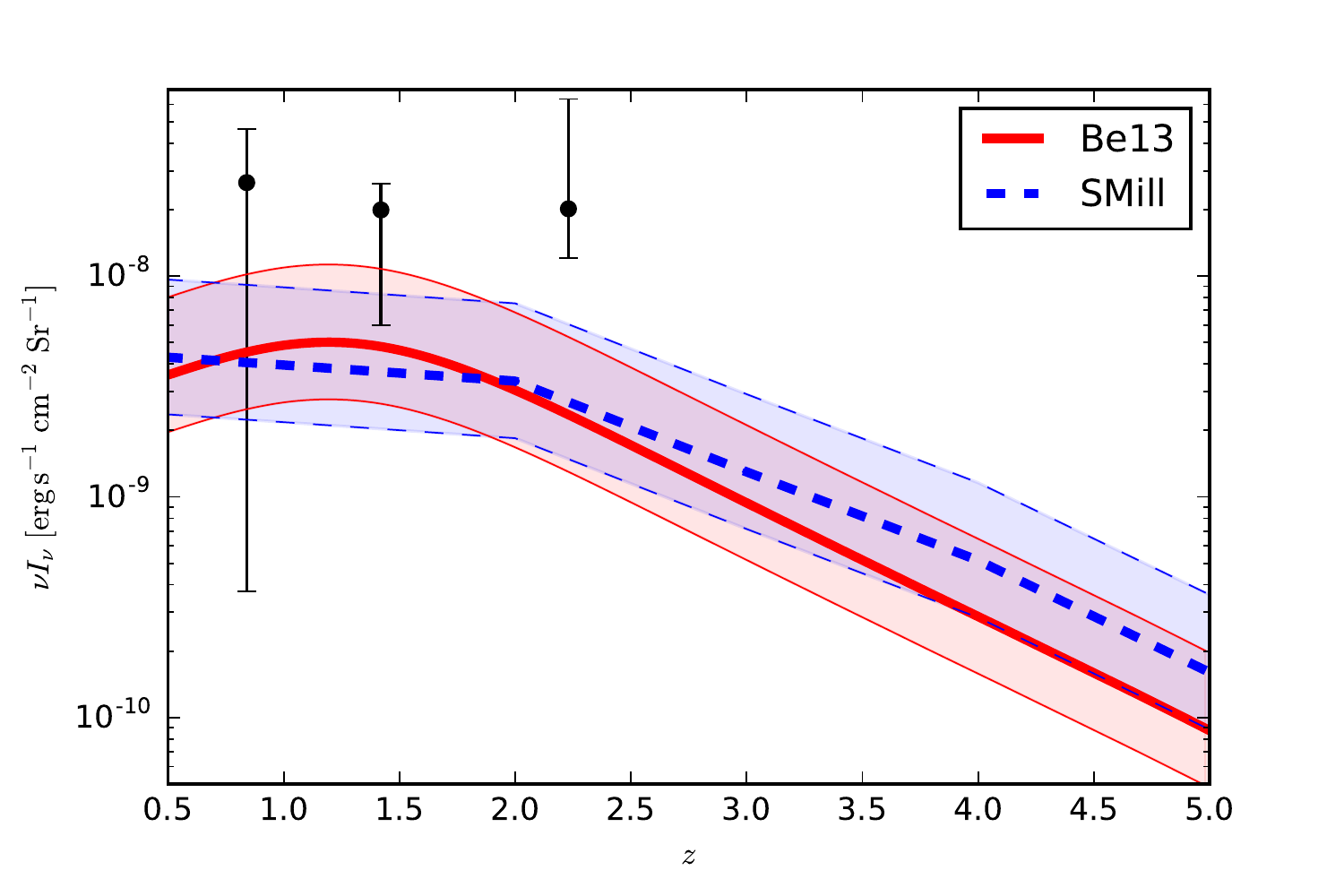}
\caption{Average OIII intensity as a function of redshift. The solid lines correspond to the estimates of the intensity using the Be SFR model \citep{Behroozi:2012iw} in red and SMill SFR model \citep{DeLucia:2006vua,Guo:2010ap} in blue. The shaded region encompass the uncertainties in $R^{\rm OIII}$, $f_{esc}^\uv$ and $E_{\rm OIII}$ and are bounded by thinner lines. The black dots were computed using the OIII luminosity functions given by \citet{Khostovan:2015aa} at $z=0.84,1.42,2.23$, and corrected with $f_{esc}^{\rm OIII}$.}
\label{fig:nuI:oiii}
\end{centering}
\end{figure}

Other optical oxygen lines worth mentioning are the OIII doublet at [500.7 nm] and [495.9 nm] with the ratio OIII[500.7 nm]/[495.9 nm]$\sim3$. These two lines are very hard to distinguish and we will therefore consider the bundle of the two as our ``estimator". We take the luminosity to be linear in the SFR and assume that $K^{\rm OIII}$ follows Eq. (\ref{eq:Kz}) with
\be \label{Roiii}
R^{\rm OIII} = \l1.3 ^{+1.2}_{-0.4}\r\times 10^{41}  ~{\rm erg/s}\,.
\ee
The total luminosity of the OIII lines was estimated by \citet{Ly:2006hx} from 197 galaxies in the redshift range 0.07 to 1.47 in the Subaru Deep Field. The UV escape fractions are the same as before and at these wavelength, and we expect an extinction of $E^{\rm OIII}=1.35$ mag. We then use Eq. (\ref{eq:savehm}) to estimate the average intensity with the same mass cutoff as for OII.
%a mass cut-off of $M_{min}=10^{12}$ M$_\odot$. 
In figure \ref{fig:nuI:oiii} the thick lines correspond to our theoretical estimate of the average OIII intensity, while the shaded area encompasses the variations of the OIII extinction, UV escape fractions and of $R^{\rm OIII}$. The OIII LF was determined from the OIII+H$\beta$ LF calibrations of \citet{Khostovan:2015aa} corrected by the extinction at these wavelengths and from which we subtracted the observed H$\beta$ LF. In the same way as for H$\beta$ studying this line is important because it is a background/foreground to be cleaned from the signal, as well as being a potential line to cross-correlate with. We can see in figure \ref{fig:nuI:oiii} that the estimates using the SFR and LF disagree. This may be caused either by the fact we are considering a lower mass cutoff using the SFR, or by the fact we are overestimating the $H\beta$ contribution for the OIII+H$\beta$ LF. Another plausible explanation concerns the metallicity of the galaxies which maybe be lower than the one assumed to estimate $R^{\rm OIII}$.

\subsubsection{Rest-frame emission in the Infrared}

The other set of relevant oxygen lines are in the far-infrared. Dust clouds around star forming regions act as re-processing bolometers and are the source of emission of OI[145$\mu$m], OIII[88$\mu$m], OI[63$\mu$m] and OIII[52$\mu$m]. \cite{2012ApJ...745..171S} gives a recipe to relate their luminosities to FIR luminosity. These were calibrated using several line luminosities of observed galaxies for which the IR continuum luminosity was available. Then, to relate the FIR luminosity to the SFR we use \citet{Kennicutt:1998zb}
\be \label{eq:L_FIR}
L_{\rm FIR} = 2.22\times 10^{43} ~\frac{SFR}{\rm M_\odot ~{\rm yr}^{-1}} \quad {\rm erg/s}\,.
\ee  
In terms of Eq.(\ref{eq:LSFR}) we have \citep{2012ApJ...745..171S}
\bea \label{Kg_osfir}
K_{\rm OI[145\mu m]}=10^{39.54\pm0.31} \,, \gamma_{\rm OI[145\mu m]}=0.89 \pm 0.06\,,\nn\\
K_{\rm OIII[88\mu m]}=10^{40.44\pm0.53} \,, \gamma_{\rm OIII[88\mu m]}=0.98 \pm 0.10\,,\nn\\
K_{\rm OI[63\mu m]}=10^{40.60\pm0.17} \,, \gamma_{\rm OI[63\mu m]}=0.98 \pm 0.03\,,\nn\\
K_{\rm OIII[52\mu m]}=10^{40.52\pm0.54} \,, \gamma_{\rm OIII[52\mu m]}=0.88 \pm 0.10\,.
\eea
Note that these are observational fits and therefore already include extinction of the lines. Later, in figure \ref{fig:nuI:comp_fir}, we show their intensities based on the previous relations and Be13 SFR model. Note that we used the same halo mass interval as for the other oxygen lines. We compared our estimates with ones using the FIR luminosity function of \citet{Bethermin:2010ea} and found that they are in good agreement. From figure \ref{fig:nuI:comp_fir} one can clearly see that some FIR oxygen lines are subdominant with respect to others while OI[63$\mu$m] and OIII[52$\mu$m] are of the same order of magnitude. Since they are not good candidates to be used as prime IM tracers, we do not show a comparison with LF estimates. Instead we only show comparisons with other FIR lines later in the paper. They nonetheless follow the dark matter distribution which, in principle, can be recovered using cross-correlations. One should also note that these FIR lines can be easily confused with each-other as well as with NIII[58$\mu$m] and CII[158$\mu$m], as is noticeable in Fig. \ref{fig:nuI:comp_fir}.

\subsection{Ionized Nitrogen}

The FIR ionized nitrogen, NII $[122\mu m]$ and NIII $[58\mu m]$, are cooling lines whose luminosities are obtained in the same way as for the FIR oxygen lines. From \citet{2012ApJ...745..171S} we have
\bea  \label{Kg_nfir}
K_{\rm NII[122\mu m]}=10^{39.83\pm0.20} \,, \gamma_{\rm NII[122\mu m]}=1.01 \pm 0.04\,,\nn\\
K_{\rm NIII[58\mu m]}=10^{40.25\pm0.55} \,, \gamma_{\rm NIII[58\mu m]}=0.78 \pm 0.10\,.
\eea
In the same way as for the FIR oxygen lines, these fits are observational and therefore do not need to be corrected for extinction. The intensity estimates are shown in figure \ref{fig:nuI:comp_fir}. As we can see they are not suitable for IM but the considerations made for FIR oxygen lines also hold for nitrogen. We still have to understand the emission of these lines not only to have further means of probing the underlying density field (using cross-correlations), but also to clean them from the signal of other lines such as CII.

\subsection{CII}

In most star forming galaxies the CII [158$\mu$m] transition provides the most efficient cooling mechanism for the gas in 
photo-dissociating regions (PDRs). This line is also emitted from ionized regions, cold atomic gas and CO dark clouds. The CII line 
is therefore often the strongest infrared emission line in galaxy spectra. This line has been proposed as a probe of the EoR \citep{Gong:2011mf,Silva:2014ira} as well as a cosmological probe for $0.5<z<1.5$ \citep{Uzgil:2014pga}. CII emission is powered by stellar UV emission and so it correlates well with the galaxy SFR. 

\citet{DeLooze:2011uw} made an observational fit to galaxy spectra at redshifts ${\rm z> 0.5}$ and determined the average CII luminosity to scale with the SFR as 
\be \label{eq:ciisfr}
L_{\rm CII} = 5.06\times 10^{40} \l\frac{SFR}{\rm M_\odot ~{\rm yr}^{-1}}\r^{1.02} \quad \rm erg/s\,.
\ee 
From Eq. (\ref{eq:LSFR}) we have $K_{\rm CII}=5.06\times 10^{40}$ and $\gamma_{\rm CII}=1.02$. This fit has a 1-$\sigma$ log uncertainty of 0.27 dex. It was based on CII line luminosity measurements and on a SFR estimated from both far-UV and ($24 \mu m$) data. 

The fit is. however, not appropriate for massive galaxies where the far-UV radiation field is very strong and so the preferential 
cooling channel is OI[63]. Also, very low mass systems are expected to be very metal poor so they should have smaller CII 
emission rates than the fit predicts. Still, we use Eq. (\ref{eq:ciisfr}) as our luminosity estimate of CII but cut off the mass integral below $M=10^{10} M_\odot$. In fact, the additional contribution from lower mass galaxies accounts for an increase in the intensity of at most $0.3\%$. The mass cutoff at $10^{10}{\rm M}_\odot$ was chosen because the observationally based relations used to convert between SFR to $L_{\rm CII}$ are consistent with emission from galaxies above $\sim 6\times10^{10} {\rm M}_\odot$. One should also stress that low mass galaxies are expected to be metal poor.

\begin{figure}
\begin{centering}  
\includegraphics[angle=0,width=\columnwidth]{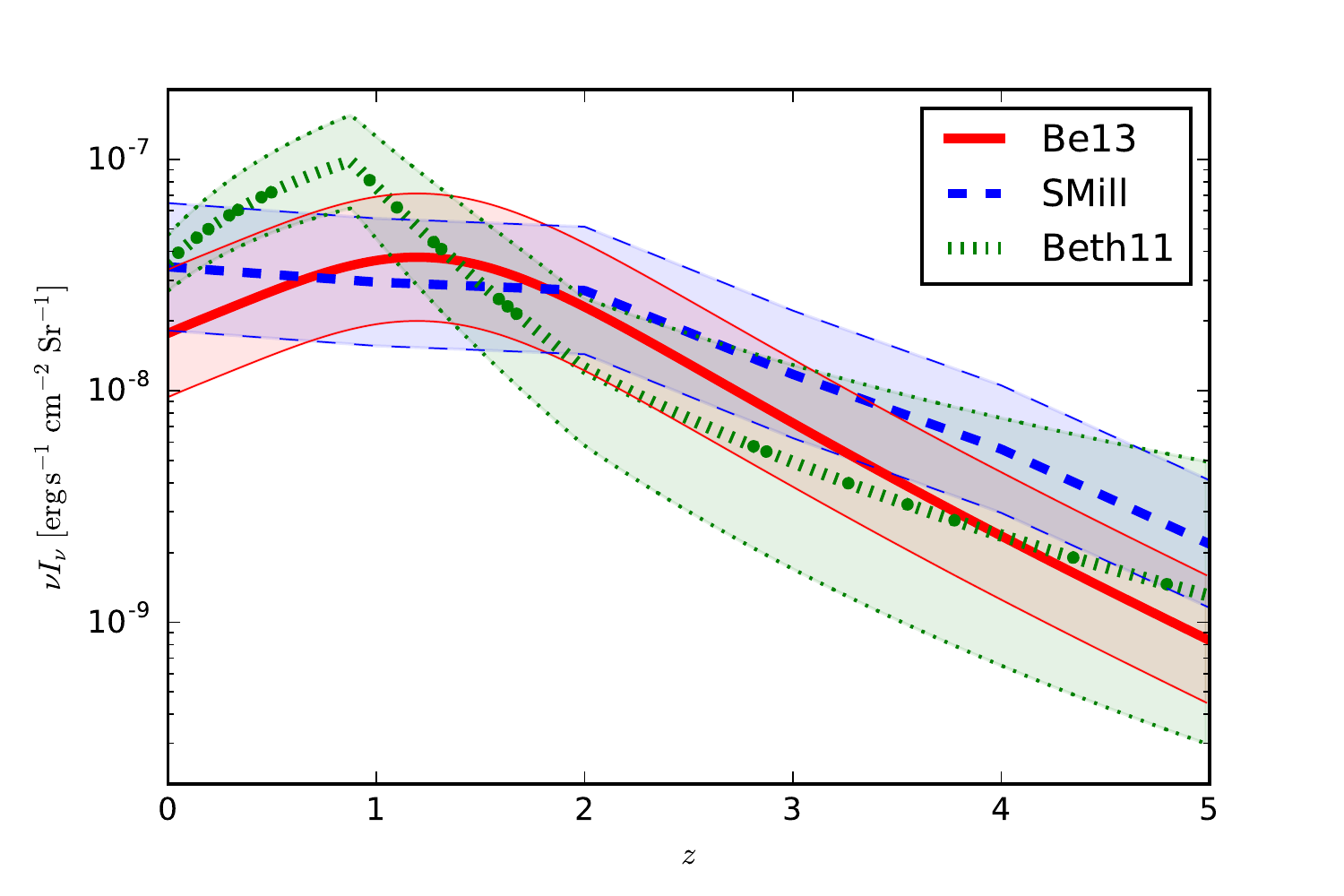}
\caption{Average CII intensity as a function of redshift using Eqs. (\ref{eq:savehm}) and (\ref{eq:ciisfr}). The thick solid red line correspond to estimates using the Be13 SFR model \citep{Behroozi:2012iw} and the thick dashed blue to the estimates obtained with SMill SFR model \citep{DeLucia:2006vua,Guo:2010ap}. The thick dotted green line is computed using Eq. (\ref{eq:ILF}) and the luminosity function given by \citet{Bethermin:2010ea}. The shaded regions correspond to uncertainties in the parameters and are bounded by thinner lines.}
\label{fig:nuI:cii}
\end{centering}
\end{figure}

In figure \ref{fig:nuI:cii} we estimate the average CII intensity as a function of redshift from Eqs. (\ref{eq:savehm}) and (\ref{eq:ciisfr}) using Be13 (solid red) and SMill (dashed blue) SFR models. The thick green dotted line is an estimate of the average CII intensity using the luminosity function given by \citet{Bethermin:2010ea} in Eq. (\ref{eq:ILF}). The shaded regions correspond to uncertainties in the estimates. We can can see that our estimates are in good agreement with the estimations using the LF. 
 
Since CII is an infrared line, intensity maps of the CII line will be contaminated by emission from other infrared 
lines, namely infrared oxygen and nitrogen lines as well as lines from CO rotational transitions at higher redshifts. Figures \ref{fig:nuI:comp_fir} and \ref{fig:nuI:comp_cos} make this point clearer. Although we didn't study the CI(1-0) and CI(2-1) lines here, they are the next subdominant lines to be considered.    

\subsection{CO}

Carbon monoxide (CO) rotational transitions are a powerful probe of the molecular gas in galaxies and of the 
astrophysical conditions in this medium. This is because the relative intensity of different transitions constrains the gas 
electron density and temperature. The excitation state of high CO transitions is correlated with the intensity of the 
radiation field exciting the molecules.% In and so in order to model CO emission the ratios of the CO luminosities 
%to the galaxy IR luminosity should be appropriate for both star forming galaxies and starbursts.

%The CO(1-0) [2600.85 $\mu$m] line is a strong molecular line and so it is commonly used as a tracer of molecular gas. Here we will mainly discuss intensity mapping of this line since the intensity of higher transitions will be considerably lower and is mostly important as a foreground line contaminant. 

\begin{figure}
\begin{centering}  
\includegraphics[angle=0,width=\columnwidth]{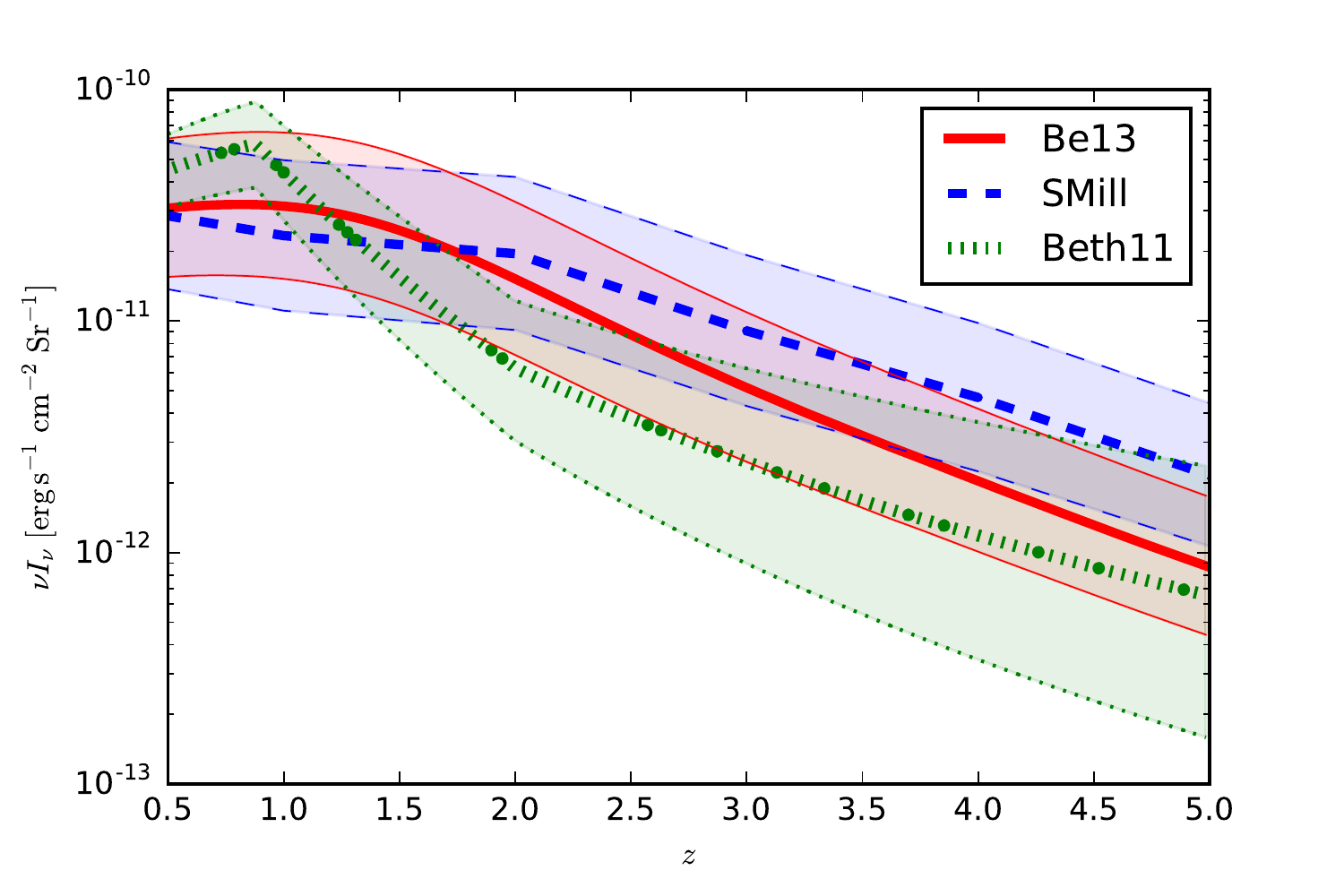}
\caption{Average CO(1-0) intensity as a function of redshift using Eqs. (\ref{eq:savehm}), (\ref{eq:L_FIR}), Eqs. (\ref{eq:lco}) and (\ref{eq:lcoprime}). The thick solid red line correspond to estimates using the Be13 SFR model \citep{Behroozi:2012iw} and the thick dashed blue to the estimates obtained with SMill SFR model \citep{DeLucia:2006vua,Guo:2010ap}. The thick dotted green line is computed using Eq. (\ref{eq:ILF}) and the luminosity function given by \citet{Bethermin:2010ea}. The shaded regions correspond to uncertainties in the calibrations of the used expressions and are bounded by thinner lines.}
\label{fig:nuI:CO10}
\end{centering}
\end{figure}

The lowest CO rotational transitions %($3\rightarrow2$ till $1\rightarrow0$) 
can be approximated by %\citep{2014ApJ...794..142G}
\be \label{eq:lcoprime}
\log_{10}\l L'_{CO}[{\rm K\, km/s\, pc^2}]\r=\alpha\log_{10}\l L_{FIR}[{\rm L_{\odot}}]\r+\beta\,.
\ee
%For starburst galaxies (SBG) one has $\alpha_{SBG}=0.99\pm0.04$ and $\beta_{SBG}=1.9\pm0.4$ \citep{2014ApJ...794..142G}
For normal star forming galaxies the CO(1-0) transition has $\alpha_{CO(1-0)}=0.81\pm0.03$ and $\beta_{CO(1-0)}=0.54\pm0.02$ \citep{Sargent:2013sxa}. The CO luminosity in erg/s can be obtained with the conversion \citep{Carilli:2013qm}
\be \label{eq:lco}
L_{CO}=1.88\times 10^{29}\l\frac{\nu_{\rm CO,rest}}{115.27 {\rm GHz}}\r^3 \frac{L'_{CO}}{\rm K\, km/s\, pc^2}\, \rm erg/s\,.
\ee
Note that $\nu_{\rm CO(1-0),rest}=115.27$ GHz. In figure \ref{fig:nuI:CO10} we present our estimates of the CO(1-0) intensity as a function of the redshift for the two SFR models considered. In addiction, we present the estimate for the intensity using the \citet{Bethermin:2010ea} IR LF. The shaded areas indicate the uncertainties in the models, showing that the different estimates broadly agree with each other.

For CO(3-2) and CO(2-1) we use the CO ratios ($R_{i1}\equiv L'_{CO(i-(i-1))}/L'_{CO(1-0)}$) from \citet{Daddi:2014mpa}, given by $R21=0.76\pm0.09$ and $R31=0.42\pm0.07$. These held similar results to the ones shown in figure \ref{fig:nuI:CO10} and are consistent with the estimates using luminosity functions. Later, in Figure \ref{fig:nuI:comp_cos} one can see how the intensity of the different CO lines compare.

For CO transitions (4-3) and higher, we assume that their luminosities follow the recent fit to the Herschel SPIRE FTS observations of 167 local normal and starburst galaxies presented in \citet{2015ApJ...810L..14L}, i.e.,
\be
\log_{10}(L'_{CO}(J)[{\rm K\, km/s\, pc^2}])=\log_{10}(L_{FIR}[{\rm L_{\odot}}])-A(J)
\ee
where $A(J=4)= 1.96 \pm 0.07$, $A(J=5)= 2.27 \pm 0.07$, $A(J=6)= 2.56 \pm 0.08$ and $A(J=7)= 2.86 \pm 0.07$.%, $A(J=8)= 3.04 \pm 0.08$, $A(J=9)= 3.20 \pm 0.09$, $A(J=10)= 3.38 \pm 0.10$, $A(J=11)= 3.56 \pm 0.11$. and $A(J=12)= 3.77 \pm 0.15$. 
 We plot all CO rotation lines up to $J=7$ in figure \ref{fig:nuI:comp_cos}.

\subsection{Other lines}

As a matter of completeness it is worth leaving a note about fainter lines which will be foreground and background contaminants. In the optical range these include other lines of the Balmer series like H$\gamma$ at 434.1nm, and stronger metal lines as NII[655.0nm], NI[658.5nm], SII[671.8nm] and SII[673.3nm]. The sodium and sulphur lines are expected to be stronger than the Balmer series ones. \citet{Uzgil:2014pga} looked at fainter lines in the FIR as SiII[35$\mu$m], SIII[33$\mu$m], SIII[19$\mu$m], NeII[13$\mu$m] and NeIII[16$\mu$m] in conjugation with FIR oxygen lines and ionized nitrogen lines. % goes to discussion:Estimates of these fainter lines could be cleaned from the signal but cross maps of different lines lines to improve the measurement of the line intensity fluctuations on the largest scales. 
These lines are generally weaker, but future studies will have to take them into account to properly determine the clean IM signal of stronger emission lines.

\section{Comparison between lines} \label{sec:line_comp}

\subsection{Measuring line intensity fluctuations} 

\begin{figure}
\begin{centering}  
\includegraphics[angle=0,width=\columnwidth]{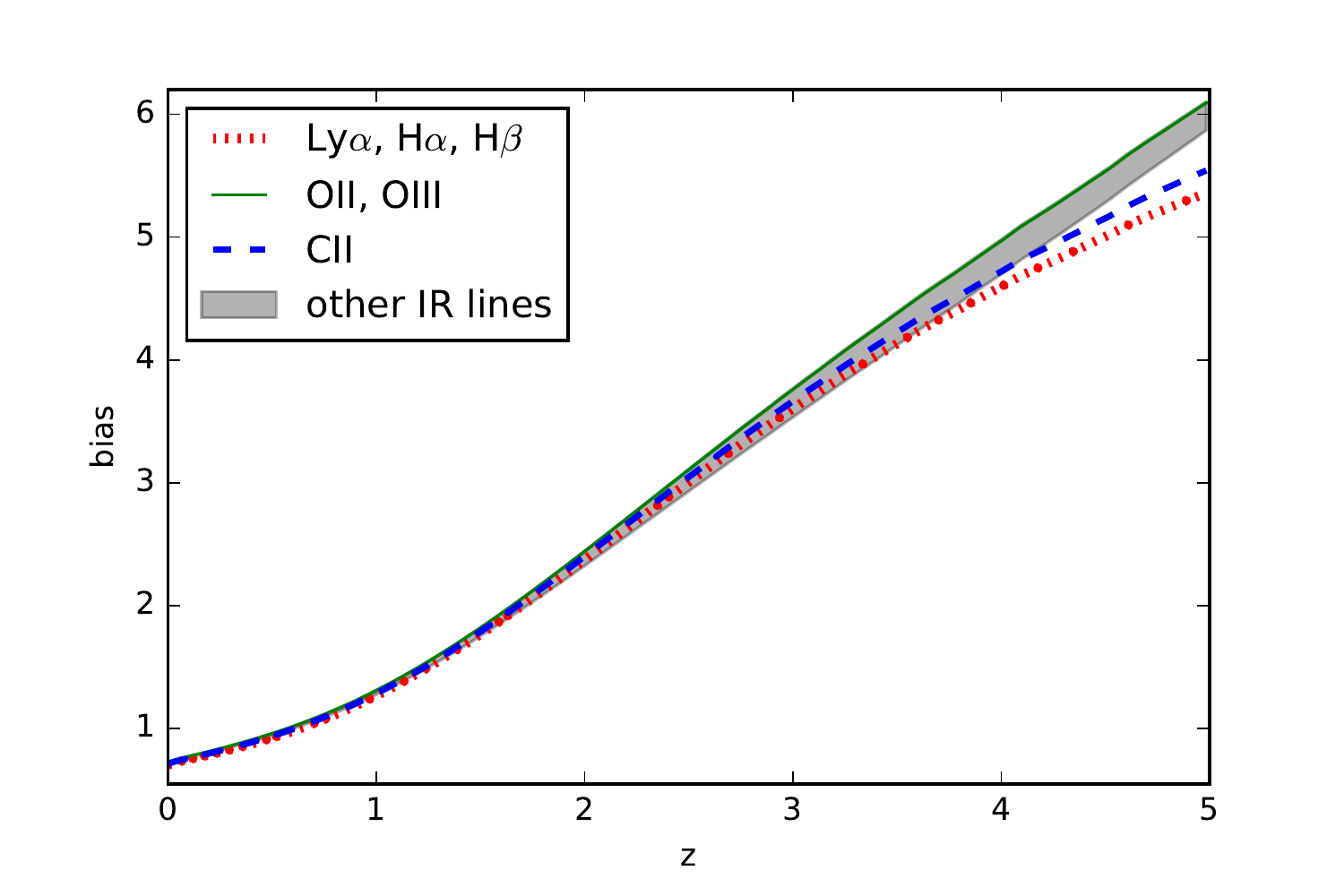}
\caption{Bias of the different lines as a function of redshift. Note that most infrared lines have similar biases hence we show them as a shaded region. It is also clear that as we increase the mass cut-off the bias of the lines increases.}
\label{fig:bias}
\end{centering}
\end{figure}

Following \citet{Visbal:2010rz}, the spatial fluctuations in the line signal linearly trace the dark matter density contrast:%. It is given by
\be
\Delta I_\nu(\theta_1,\theta_2,z)\equiv I_\nu(\theta_1,\theta_2,z)-\bar I_\nu(z)= \bar I_\nu(z) \Delta\l \bf x ,z\r\,,
\ee
where $I_\nu$ is the point dependent signal and $\bar I_\nu$ is the average line signal in a redshift bin. In the simplest scenario, neglecting redshift-space distortions, lensing and relativistic corrections, $\Delta\l \bf x ,z\r=\bar b(z) \delta\l \bf x ,z\r$, where $\bar b$ is the luminosity weighted bias of the emission line and $\delta\l \bf x \r$ is the underlying dark matter density contrast. The 3D power spectrum of the emission line is simply given by
\be
P_I\l z, k\r= \bar I_\nu^2\l z\r  \bar b^2\l z\r~P_{CDM}\l z, k \r\,,
\ee
where $P_{CDM}$ is the dark matter power spectrum. Lastly, we should also estimate the shot noise contribution to each line, which is usually negligible. It is given by \citep{2011ApJ...740L..20G}
\be \label{eq:pshotline}
P_{shot} (z)=\int_{M_{min}}^{M_{max}} \d {\rm M}~\hmf \l\frac{L(M,z)}{4\uppi D_L^2}~\tilde y D_{A}^2\r^2 \,.
\ee
Note that contrary to threshold surveys, this shot noise is independent of the survey/instrument specifications. 
The luminosity weighted bias is given by
\be
\label{eq:lumbias}
\bar b \l z\r \equiv\frac{\int^{M_{max}}_{M_{min}}{\rm dM} ~b\l M,z\r L(M,z) ~\hmf }{\int^{M{max}}_{M_{min}}{\rm dM}~ L(M,z) ~\hmf }\,,
\ee
where $b\l M,z\r$ is the halo bias, $L(M,z)$ is the line luminosity and $dN/dM$ is the halo mass function. Using this definition and taking Eq. (\ref{eq:LSFR}), one can write $\bar b$ in terms of the SFR: 
\be
\label{eq:bias:simple}
\bar b \l z,\gamma\r =\frac{\int^{M{max}}_{M_{min}}{\rm dM}~ b\l M,z\r \l SFR(M,z)\r^\gamma ~\hmf }{\int^{M{max}}_{M_{min}}{\rm dM} ~\l SFR(M,z)\r^\gamma ~\hmf }\,.
\ee
Hence, the bias of each emission line depends on the mass cut-off and the value of $\gamma$. Without any other contribution, or dependence on the mass, one expects $\lya$, $\ha$ and H$\beta$ to have the same bias. The bias difference between the hydrogen lines and the optical OII and OIII doublet comes from the lower halo mass cut-off. On the other hand the bias differences between CII, the lowest CO lines ($J<3$), the infrared ionized oxygen and nitrogen lines come from the fact that they have a non-linear dependence on the SFR (i.e., $\gamma \neq1$). One can see this behavior in Fig. \ref{fig:bias}. Note that most FIR lines have similar biases which we only represent as a shaded area. We should note that since the CO rotation lines for $J\geq4$ are linear in the SFR and, to first approximation, the full range of DM halos emit these lines, we expect the bias of CO($J\geq4$) to be similar to the Hydrogen emission lines. All these bias we computed using Be13 SFR model in Eq. (\ref{eq:bias:simple}).

\subsection{Range of lines dominance and contaminants}\label{subsec:domline}

\begin{figure}
\begin{centering}  
\includegraphics[angle=0,width=\columnwidth]{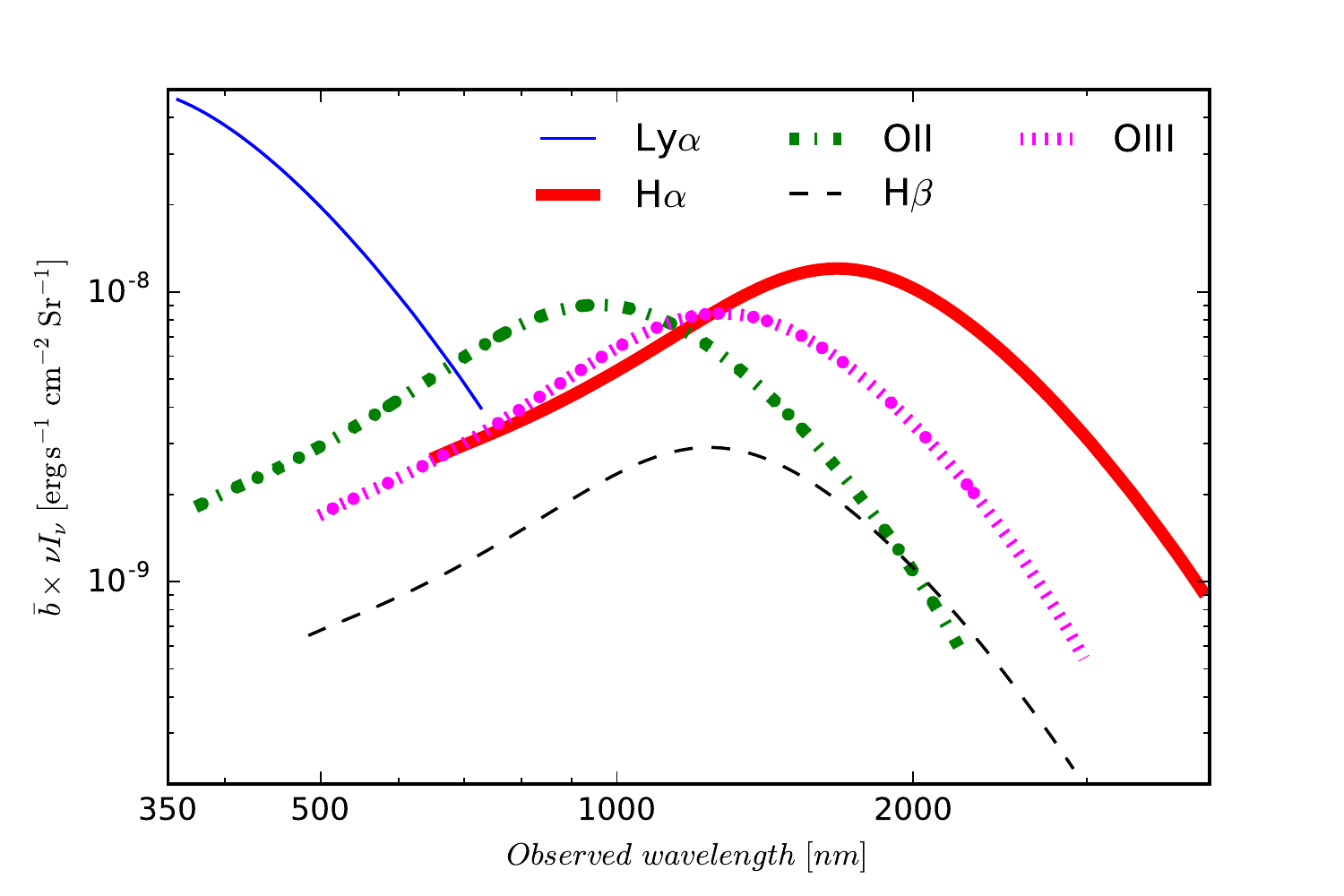}
\caption{{\bf UV/Optical/NIR:} Estimates of the product $b\times\nu I_\nu$ of $\lya$, $\ha$, H$\beta$, OII and the OIII doublet as a function of the observed wavelength. Due to Earth's observational constraints and the difficulties in UV observations we only consider observed emission from the near UV till NIR. All plotted lines are for the redshift interval $z=0-5$, except $\lya$ where we cut below $z\sim1.9$.}
\label{fig:nuI:comp_opt}
\end{centering}
\end{figure}

In Fig. \ref{fig:nuI:comp_opt} we plot the product of the bias with the estimated average intensity of the UV and optical lines studied previously. $\lya$, $\ha$, H$\beta$, OII and the OIII doublet are plotted as a function of the observed wavelength up to redshift 5.
% in the optical and infrared . 
%The ranges in the plot consider both models Be12 and SMill to determine the highest and lowest possible emission strength. 
Note that each line was estimated using the Be13 model. As expected, the emission from $\lya$ is expected to be highest; this line is therefore a very strong candidate to use for IM techniques. One can also see that OII dominates for a narrow wavelength range, while $\ha$ clearly dominates in the NIR. For $z>5$ $\lya$ will be a background contaminant of OII, while for $z\lesssim0.2$ $\ha$ will be a foreground. Similarly, OII will be a foreground for $\lya$ and will give background contamination to $\ha$. As stated before, the H$\beta$ and the OIII doublet will mainly be contaminants or lines to cross-correlate with. One should bear in mind that although some lines are the dominant contributor in a particular wavelength regime, the power spectrum of subdominant lines may shoot these up locally. We will not address this here and leave it for future work since such issues are highly dependent on the chosen line. %   that need to be addressed. %Still they carry cosmological information that, to our knowledge, can only be extracted using cross-correlations, if at all. %\tbe{develop this idea of autocorrelate with wavelength displaced, i.e., same redshift.}

\begin{figure}
\begin{centering}  
\includegraphics[angle=0,width=\columnwidth]{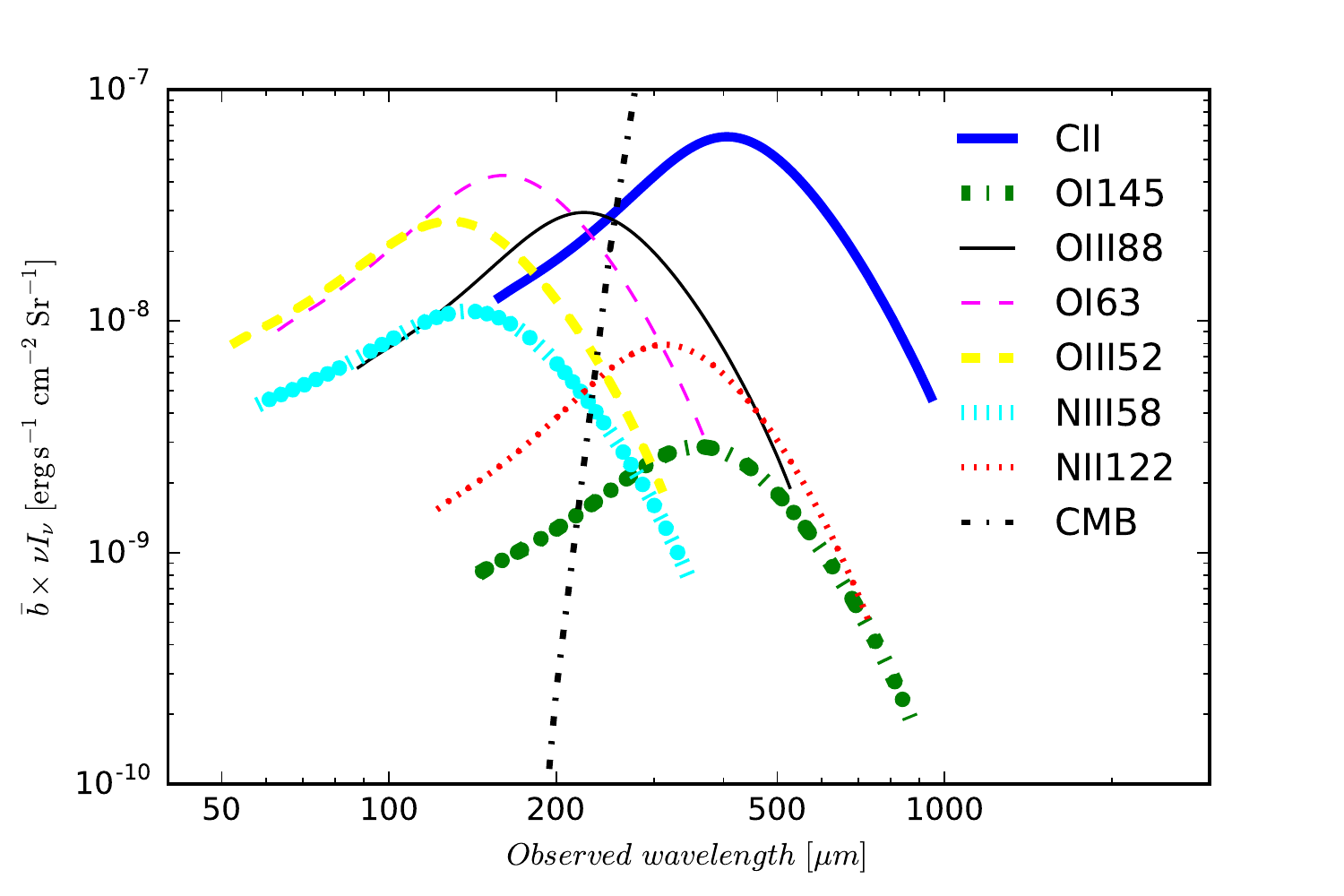}
\caption{{\bf FIR:} Estimates of the product $b\times\nu I_\nu$ of CII and the other Oxygen and Nitrogen FIR emission lines as a function of the observed wavelength. We also plot the CMB intensity for comparison. All plotted lines are for the redshift interval $z=0-5$.}
\label{fig:nuI:comp_fir}
\end{centering}
\end{figure}

From Fig. \ref{fig:nuI:comp_fir} we can clearly see that CII is the dominant FIR emission line and has the potential to be used for IM after cleaning the CMB signal from the map. Still, at lower redshifts it becomes subdominant and no other line becomes clearly dominant. In contrast, the oxygen and nitrogen lines are not the most promising lines for IM. They easily contaminate the signal of each other and CII at lower redshifts. Additionally, CI, Si and Ne lines, which have not been included, will further contaminate the signal \citep{Uzgil:2014pga}. 

\begin{figure}
\begin{centering}  
\includegraphics[angle=0,width=\columnwidth]{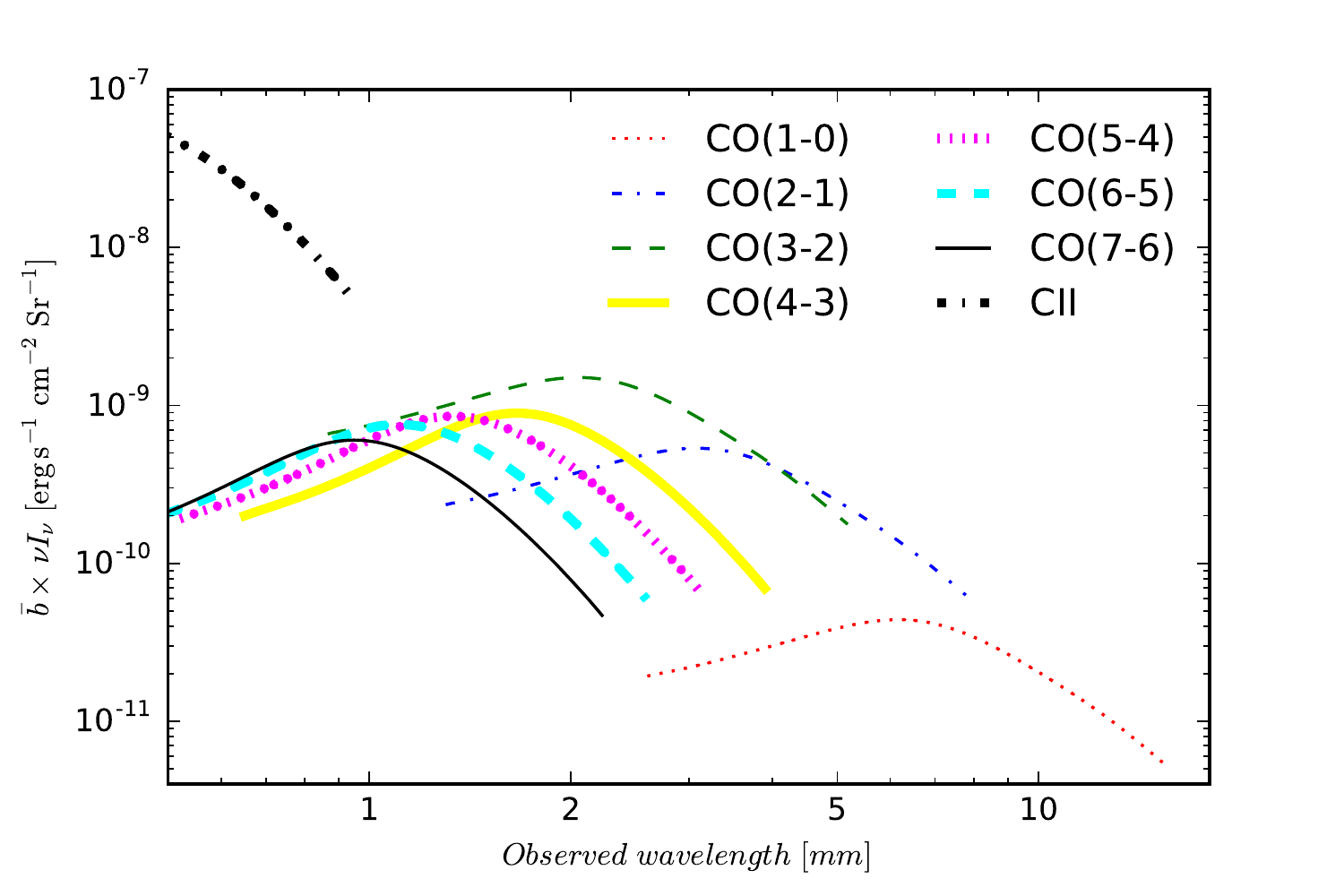}
\caption{{\bf COs:} Estimates of the product $b\times \nu I_\nu$ CO rotational transitions as a function of the observed wavelength. All plotted CO lines are for the redshift interval $z=0-5$. We also plot the CII emission for comparison.}
\label{fig:nuI:comp_cos}
\end{centering}
\end{figure}

In figure \ref{fig:nuI:comp_cos} we compare the bias times the intensity of several rotational transitions of CO. We also plot the high redshift emission of CII as comparison. We do not show the CMB emission since it will be just a background at a fixed temperature. One can see that the lowest three lines could be used for IM at intermediate to high redshifts. In the case of CO(1-0) contamination will mainly come from CO(2-1) at redshifts below $\sim6.2$. Both CO(2-1) and CO(3-2) will be contaminated by higher J rotation lines at lower redshift and therefore cannot be used for IM in these regimes. Also, one can see that CO(3-2) will be subdominant with respect to CO(2-1) at $z\gtrsim 3.5$ but at intermidiate redshifts one can still use it as an IM probe. 

\begin{table}
\centering
\caption{Observational wavelength range and corresponding redshift range (up to $z=5$) for which lines dominate the signal based on our estimates using Be13 model \citep{Behroozi:2012iw} for the SFR. These values are to be read as indication where these lines dominate although a wide range can still be used using masking and cleaning methods.}
\begin{tabular}{ l c c c }
\hline\hline
Emission & Wavelength & Redshift & Spectral\\
line& range& range&class\\
\hline
Ly$\alpha$ &122-679 nm & 0.0-4.58 & UV/Optical\\
OII &679-1131 nm &0.82-2.03&Optical/NIF\\
%\tbd{OIII &1131-1248 nm &0.9-2.1&Optical/NIF}\\
H$\alpha$ &1.25-3.94 $\mu$m &0.9-5&NIF\\
CII &249-948 $\mu$m& 0.57-5& FIR\\
CO(3-2)  &1.24-4.06 mm & 0.43-3.69 &Radio(Millimetre)\\
CO(2-1)  &4.06-7.8 mm & 2.12-5 &Radio(Millimetre)\\
CO(1-0)  &8.48-15.6 mm & 2.60-5 & Radio(Millimetre)\\
  \hline
\end{tabular}
\label{table:IMlines}
\end{table}

Figures \ref{fig:nuI:comp_opt}, \ref{fig:nuI:comp_fir} and \ref{fig:nuI:comp_cos} merely provide an indication about which lines dominate the observed intensity and where in the spectrum. Other effects, such the power spectrum of very low redshift lines, may still create high contamination (although one expects to be able to mask it). Nonetheless we summarize in table \ref{table:IMlines} the potential intensity mapping lines and the wavelength range where their intensity is dominant up to redshift 5. We also state which redshifts could be probed by these lines. These fall in a wide range of the electromagnetic spectrum, and some lie outside the observable atmospheric windows meaning space experiments {are required}. In summary, one can say that the best candidates to be used for cosmological IM surveys at lower-to-intermediate redshifts are $\lya$, OII, $\ha$, CII and the lower CO transitions.

\section{Surveys} \label{sec:Surveys}

An IM experiment requires one to be able to separate the incoming light from a field into wavelength (and therefore redshift) bins. For ground-based telescopes one can either use low-resolution spectroscopy with narrow band filters, Fabry-P\'erot filters or Integral Field Units (IFUs) for high-resolution spectroscopy. NIR lines can only be observed from space. The FIR and sub/millimeter offer more possibilities with dish experiments and optical-like settings at high altitudes.% can offer ground based surveys.    
% One of the telescopes that takes advantage of such atmospheric windows in the infrared is the ground telescope Subaru [http://subarutelescope.org], but for a sky survey in a wide range of IR wavelengths one requires a space telescope. OII will only be dominant for Optical and NIR wavelengths and therefore ground based telescopes equipped with IFU's can perform (within their time constrains) OII IM surveys. The FIR lines of CII and CO require groundbase single dish experiments for IM surveys. 
We will discuss the possibility of measuring the power spectrum with these lines for current and proposed experiments as well as feasible setups.  

\subsection{Error estimation}

In this paper we will only focus on the detectability of the 3D power spectra of different emission lines taking into account instrumental and shot noise. % can be modeled in a straight forward fashion. 
In this section, we will neglect the uncertainties due to the parameters used to estimate the intensity, and uncertainties due to contamination from other lines, foregrounds and backgrounds. 

For an experiment with sensitivity $\sigma_N$, the noise power spectrum is given by
\be \label{eq:pnoise}
P_N=\sigma_N^2\times V_{pixel}\,,
\ee
where $V_{pixel}$ is the comoving volume corresponding to the redshift and angular resolution of the considered experiment. The simplest estimate of the error in measuring $P(k)$ is
\be
\Delta P (k_j)\simeq\frac{P_T (k_j)}{\sqrt{N_k (k_j)}}\,,
\ee
where $P_T=P_S+P_N+P_{shot}$ is the total power in a scale $k_j$, while $N_k$ is the number of accessible modes at a scale $k$. For a 3D survey $N_k(k_j)=k_j^2\Delta k V_{sample}/2\uppi^2$, where $V_{sample}$ is the comoving volume of the survey and $\Delta k$ is the chosen $k$-bins. This approach is only valid within a regime $k\in[k_{min},k_{max}]$ where $k_{min}\sim 2\uppi/L$ is given by the smallest side of the sample volume and $k_{max}\sim 2\uppi/\Delta L$ by the biggest side of the resolution pixel. Outside this range the k-space is approximately 2D, i.e, $N_k(k_j<k_{min})=k_j\Delta k S_{sample}/2\uppi$. For a single scale (bin j) the signal-to-noise is given by $SNR(k_j)=P_S(k_j)/\Delta P(k_j)$. Adding up all the signal-to-noise, for a given experiment one has
\be
SNR^2=\sum_j \l\frac{P_S(k_j)}{\Delta P(k_j)}\r^2\,.
\ee

\subsection{$\lya$ IM}

An experiment using IFUs is the HETDEX - Hobby-Eberly-Telescope Dark Energy eXperiment \citep[www.hetdex.org]{Hill:2008mv}, a 3-year survey designed to see approximately 0.8 million Lyman-$\alpha$ emitting galaxies (LAEs) in the redshift range $z\sim$1.9 - 3.5 covering 300 $\deg^2$ with a filling factor of $1/4.5$. This will be achieved using the instrument VIRUS (Visible Integral-Field Replicable Unit Spectrograph) which is composed of 150 wide-field IFUs, each with 224 optical fibers. Pairs of IFUs are built as single units so it will observe 33600 pixels, obtaining a spectrum for each pixel. HET has a field of view (FoV) of 22 arcminutes in diameter while VIRUS will only provide a coverage of around $1/4.5$ of the full FoV. Each fiber has a diameter of $1.5"$ but there will be a dithering pattern that will effectively give an angular resolution of $\delta\Omega_{pixel}\simeq9.05$ arcsec$^2\simeq2.13\times10^{-10}$Sr. A single field will be observed for 1200s using 3 separated dithering exposures, giving a total survey time of 1200h (assuming 140 observation nights over 3 years). VIRUS will have a wavelength coverage from 3500 to 5500 $\mathring A$ with a spectral resolution of $\la/\Delta\la= 800$ or an average $6.4 \mathring A$ wavelength resolution. The quoted line sensitivity for 20 minutes of integration time at $z=2.1$ is $1.28\times 10^{-17}$ erg/s/cm$^2$ or $\sigma\l\nu I_\nu\r=6.02\times 10^{-8}$ erg/s/cm$^2$/Sr.%$9.5,3.9,3.4$ and $3.5\times 10^{-17}$ erg/s/cm$^2$ at $\la_O=3500,4250,4850$ and $5500 \mathring A$ respectively. Taking the later as our reference we estimate that the flux sensitivity per solid angle to be $\sigma_{\nu I_\nu}=1.64\times 10^{-7}$ erg/s/cm$^2$/Sr.

An intensity mapping experiment with VIRUS will not need to use the full resolution of the experiment since we can consider larger IM pixels and wavelength bins. By doing so, one can reduce the experimental noise and increase the detectability of the signal.  As an example, the given resolution of $6.4 \mathring A$ in the range $3500$-$5500 \mathring A$ gives around 300 redshift bins of size $\Delta z\sim 0.005$. Such redshift resolution is unnecessary for cosmological studies as well as potentially introducing errors due to the emission line profile of Lyman-$\alpha$. For IM the width of the bin needs to be bigger than the observed FWHM of $\lya$ line. The authors \citet{Yamada:2012kn} find a rest frame $\lya$ FWHM which is smaller than $1\mathring A$. Even taking a conservative approach, at redshift 3 one only expects the observed FWHM to be $\sim 4 \mathring A$. Similarly, small angular scales are not good probes of the cosmology since they are polluted by uncertainties in the clustering of matter in dark matter halos. %As an example, the size of the pixel would correspond to $\ell_{max}\sim 2.2\times10^{5}$ which is unreasonably high. 
Although changing the pixel size could increase the sensitivity in the pixel, this does not alter the noise power spectrum (Eq. \ref{eq:pnoise}) since the decrease in the sensitivity is cancelled by the increase in the pixel volume. The only variable that one could change in HETDEX is the integration time. We can then rewrite the instrumental noise for IM as 
\be
\sigma^{\rm HETDEX}_N=\sigma_{\nu I_\nu}\sqrt{\frac{20~{\rm min}}{\delta t}}\,,
\ee
where %$\Omega_{pixel}$ is the new pixel angular size, 
$\delta t$ the new integration time per pointing. % and $\delta \lambda$ the new wavelength binning. 
One should note that for a 20 min integration, the sensitivity is up to a order of magnitude higher that the expected signal (compare with figure \ref{fig:nuI:lya}). 

\begin{figure}
\begin{centering}  
\includegraphics[angle=0,width=\columnwidth]{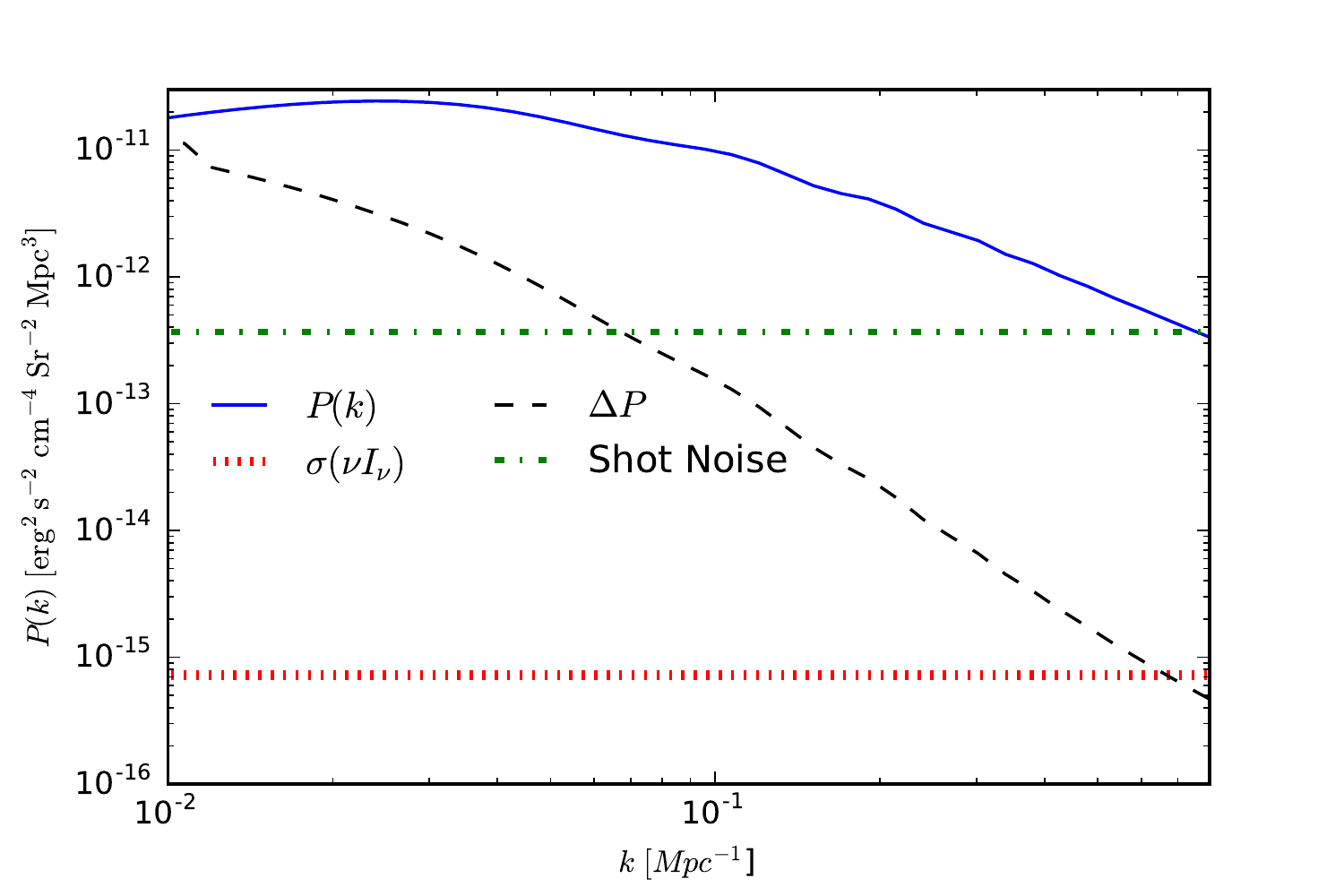}
\caption{Estimated power for $\lya$ IM (solid blue), Hetdex noise (thick dotted red), $\lya$ shot noise (dashed-dotted green) and the error in estimating the power spectrum (dashed black) at $z=2.1$.}
\label{fig:pk:lya}
\end{centering}
\end{figure}

\begin{figure}
\begin{centering}  
\includegraphics[angle=0,width=\columnwidth]{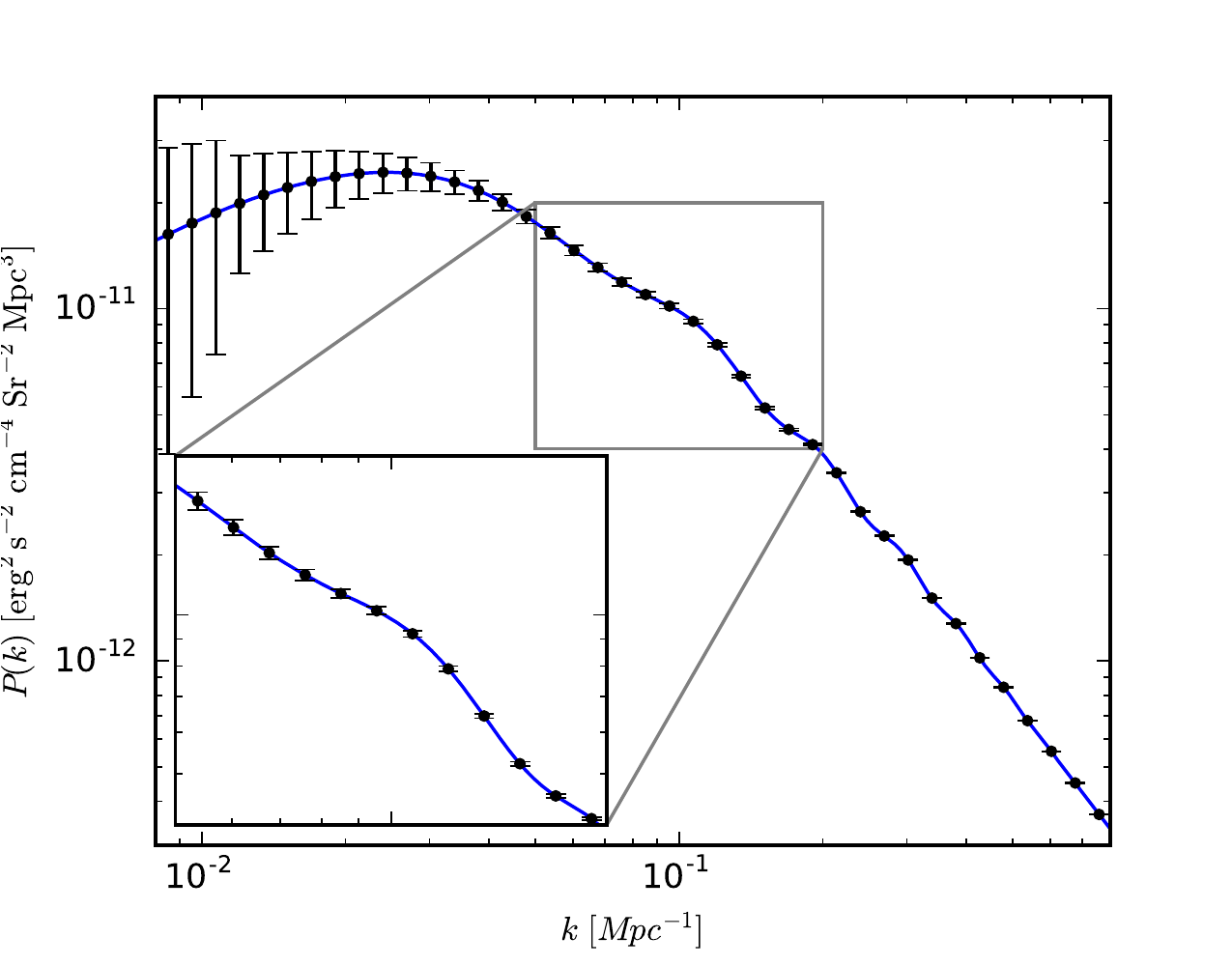}
\caption{$\lya$ IM Power Spectrum at $z=2.1$  with forecasted error bar for HETDEX.}
\label{fig:errpk:lya}
\end{centering}
\end{figure}

Note that $\delta$t is just the ratio between the total observation time $t_{TOT}$ and the number of pointings $N_p$, where $N_p$ is the ratio between the surveyed area $S_{Area}$ and FoV$_{VIRUS}$. If one takes into account the 1 minute of overheads, the integration time is $\delta t=t_{TOT}[{\rm min}]/N_{p}-1$min. The HETDEX quoted numbers point to an integration time of 20 mins for roughly 3500 pointings. Unfortunately this is for a filling factor of 4.5. If we keep the sparse sampling of the HETDEX survey then both the instrumental and shot noise need to increase by the same amount as the filling factor \citep{Chiang:2013ksa}. 

In figure \ref{fig:pk:lya} we show the power spectrum of $\lya$ IM (solid blue), the power coming from HETDEX instrumental noise (dotted red), $\lya$ shot noise (dot-dashed green) and $\Delta P_{\lya}$ (dashed black) at $z=2.1$. We assume a sparse coverage of $300\deg^2$ and a sample with $\Delta z=0.4$ around $z=2.1$. One can see that HETDEX is so sensitive that the instrumental noise is subdominant with respect to $\lya$ shot noise (although this is dependent on assumptions), while both are orders of magnitude lower than the power spectrum \emph{per se}. This becomes clearer when plotting the power spectrum with the estimated error bars, as in figure \ref{fig:errpk:lya}. Not only would one measure the $\lya$ power spectrum very well but one would also have good enough statistics to resolve the wiggles from the Baryonic Acoustic Oscillations (BAO). $\lya$ IM is therefore a good candidate to study further including all models for background continuum emission, foregrounds and contaminants. We will leave this to a future paper.

HETDEX is primarily a galaxy survey, but can be used as a $\lya$ intensity mapper as we have pointed out. We have also shown that $\lya$ intensity shot noise is much higher than the HETDEX instrumental noise, if it is used as an intensity mapping experiment. One should note that $\lya$ intensity shot noise is intrinsic to the line (although it is affected by the sampling coverage). Hence, as an IM survey HETDEX does not gain anything from longer integration times. In fact, this causes the survey to be sparse, thus increasing the shot noise, and reducing the area covered, i.e., increasing the error bars on the power spectrum due to a lower number of available $k$-modes. Therefore, one concludes that if HETDEX was an IM experiment one ought to consider larger survey areas with less time per pointing in order to maximize the signal-to-noise ratio.

\subsection{$\ha$ IM} \label{subsec:ha}

$\ha$ is an optical line at rest but will be observed in the infrared for $z\gtrsim 0.06$. The Earth's observational window in the infrared is reduced to a minor collection of narrow observational gaps in the NIR. Hence we need to make space observations, such as Euclid \citep[http://www.euclid-ec.org]{Amendola:2012ys} which will use H$\alpha$ emission for its galaxy survey.

Here we will focus on the planned space telescope \emph{SPHEREx} \citep[http://spherex.caltech.edu]{Dore:2014cca} and revisit their IM mode. SPHEREx is an all-sky space telescope in the NIR having four linear variable filters. Although it covers the full sky, only $\sim 7000\deg^2$ will be of any cosmological use. Its spectral resolution is $\la/\Delta\la= 41.5$ for $0.75<\la<4.1\mu$m and $\la/\Delta\la= 150$ for $4.1<\la<4.8\mu$m. The instrument has %FoV of $3.5\deg\times7.5\deg$ with 
a pixel size of $6.2"\times6.2"=9.03\times 10^{-10}$Sr. % Every single day SPHEREx will swap $1\deg$ of RA of the sky, observing the all sky in 6 months. In fact, due to the instruments configurations, 7 days are needed for a full degree in the sky to be seen in its complete wavelength range. Although one of the FoV sides has a $7.5\deg$ width, it will not be observing at the same wavelengths. 
 The $1\sigma$ flux sensitivity depends on the wavelength bin in consideration but remains within the same order of magnitude. For $\ha$ at $z=1.9$ we will take $\sigma\l\nu I_\nu\r\sim 1\times 10^{-6}$ erg/s/cm$^2$/Sr. 

\begin{figure}
\begin{centering}  
\includegraphics[angle=0,width=\columnwidth]{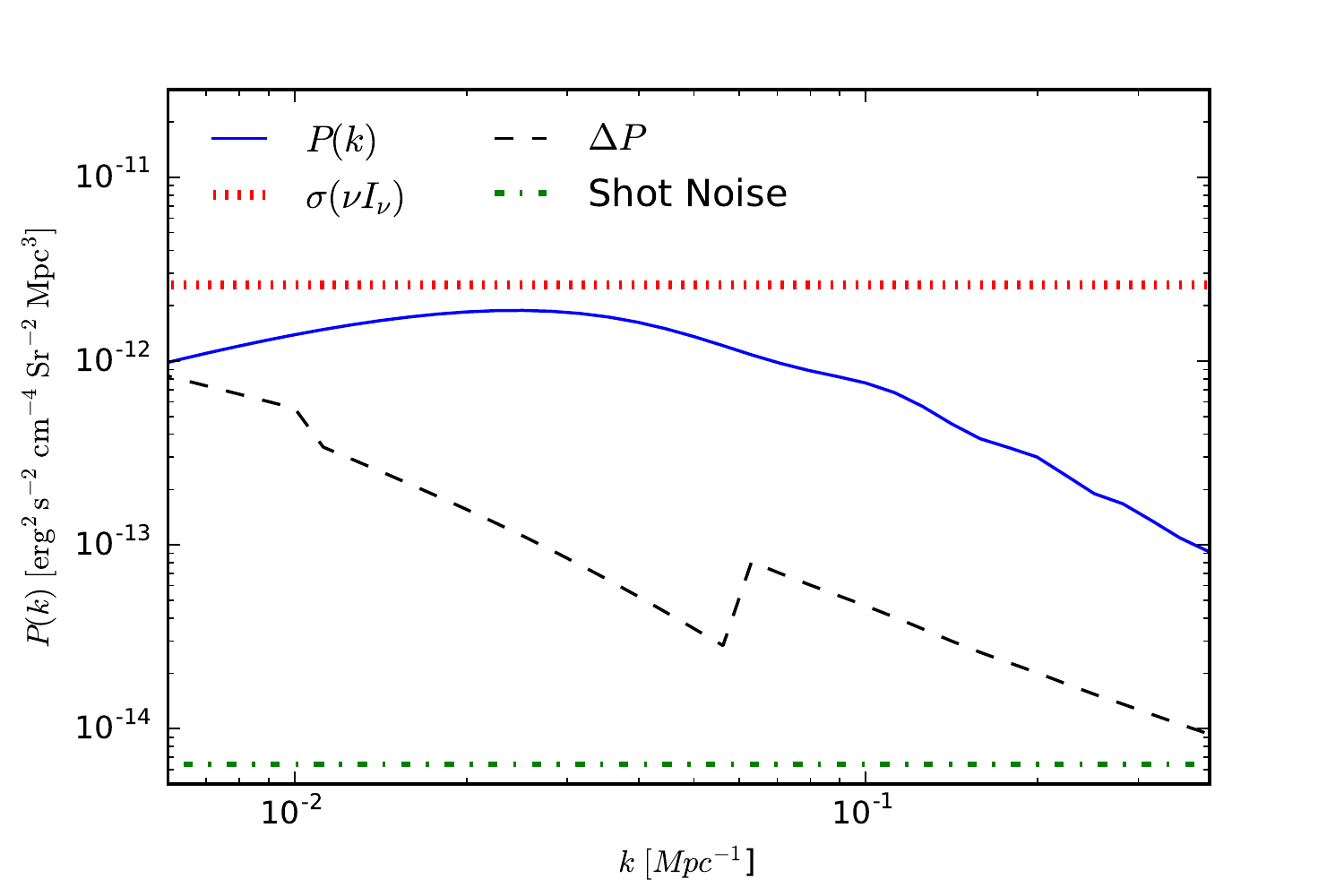}
\caption{Estimated power for $\ha$ IM (solid blue), SPHEREx noise (thick dotted red), $\ha$ shot noise (dashed-dotted green) and the error in estimating the power spectrum (dashed black) at $z=1.9$.}
\label{fig:pk:ha}
\end{centering}
\end{figure}

\begin{figure}
\begin{centering}  
\includegraphics[angle=0,width=\columnwidth]{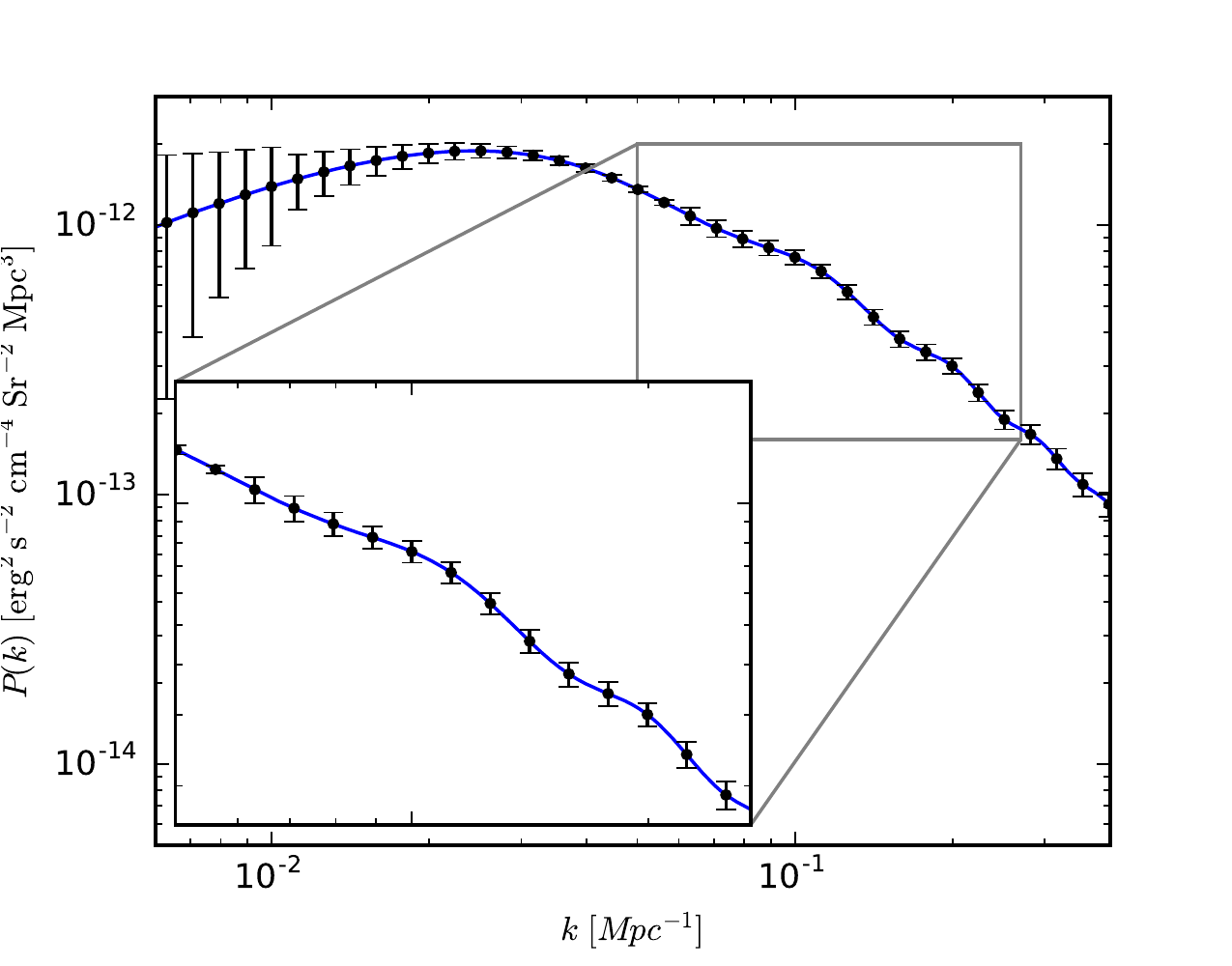}
\caption{$\ha$ IM Power Spectrum at $z=1.9$ with forecasted error bar for SPHEREx.}
\label{fig:errpk:ha}
\end{centering}
\end{figure}

SPHEREx is a space survey so we cannot change the integration time to improve the sensitivity of the experiment. The size of the pixel is only relevant for the maximum $k$ that is accessible. In fact, changing the pixel size will not alter the noise power as we saw before. As a thought experiment let us start by analyzing how well we would observe the power spectrum at $z=1.9$ with a sample size of $\Delta z=0.4$. %($\delta \lambda=0.13\mu$m). 
In figure \ref{fig:pk:ha} we show in the blue solid line the $\ha$ power spectrum at $z=1.9$, in dotted red the instrumental noise power spectrum, in dot-dashed green the $\ha$ shot noise and in the black dashed line the error in measuring the power spectrum. The number of modes mainly comes from the 2D information encoded in the surveyed area. We can see that we should be able to measure the power spectrum on large scales with SPHEREx, as well as the BAO. This becomes clearer in Fig. \ref{fig:errpk:ha}.

A more futuristic experiment is the Cosmic Dawn Intensity Mapper \citep{Cooray:2016hro}. It would cover a part of the spectrum slightly broader than SPHEREx ($\lambda=0.7-7\mu$m) and would have a flux sensitivity 30-50 times higher than SPHEREx. This would have a clear impact on the error bars of the power spectrum, especially at larger scales. Although Cosmic Dawn Intensity Mapper was proposed to study EoR it can, and should, have a commensal $\ha$ (and OII) IM survey at intermediate-to-low redshifts.

\subsection{OII IM}

\begin{figure}
\begin{centering}  
\includegraphics[angle=0,width=\columnwidth]{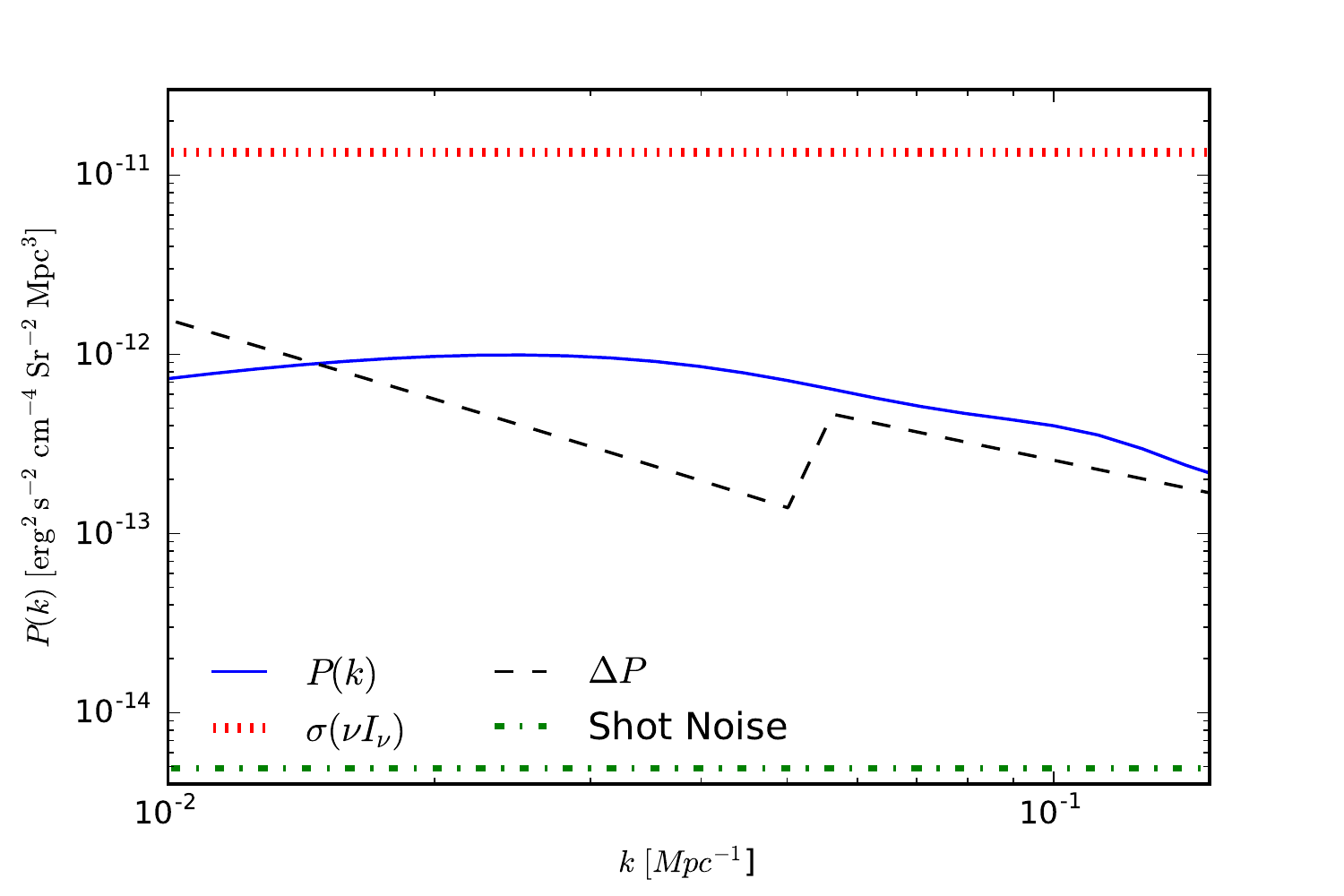}
\caption{Estimated power for OII IM (solid blue), SPHEREx noise (thick dotted red), OII shot noise (dashed-dotted green) and the error in estimating the power spectrum (dashed black) at $z=1.2$.}
\label{fig:pk:oii}
\end{centering}
\end{figure}

\begin{figure}
\begin{centering}  
\includegraphics[angle=0,width=\columnwidth]{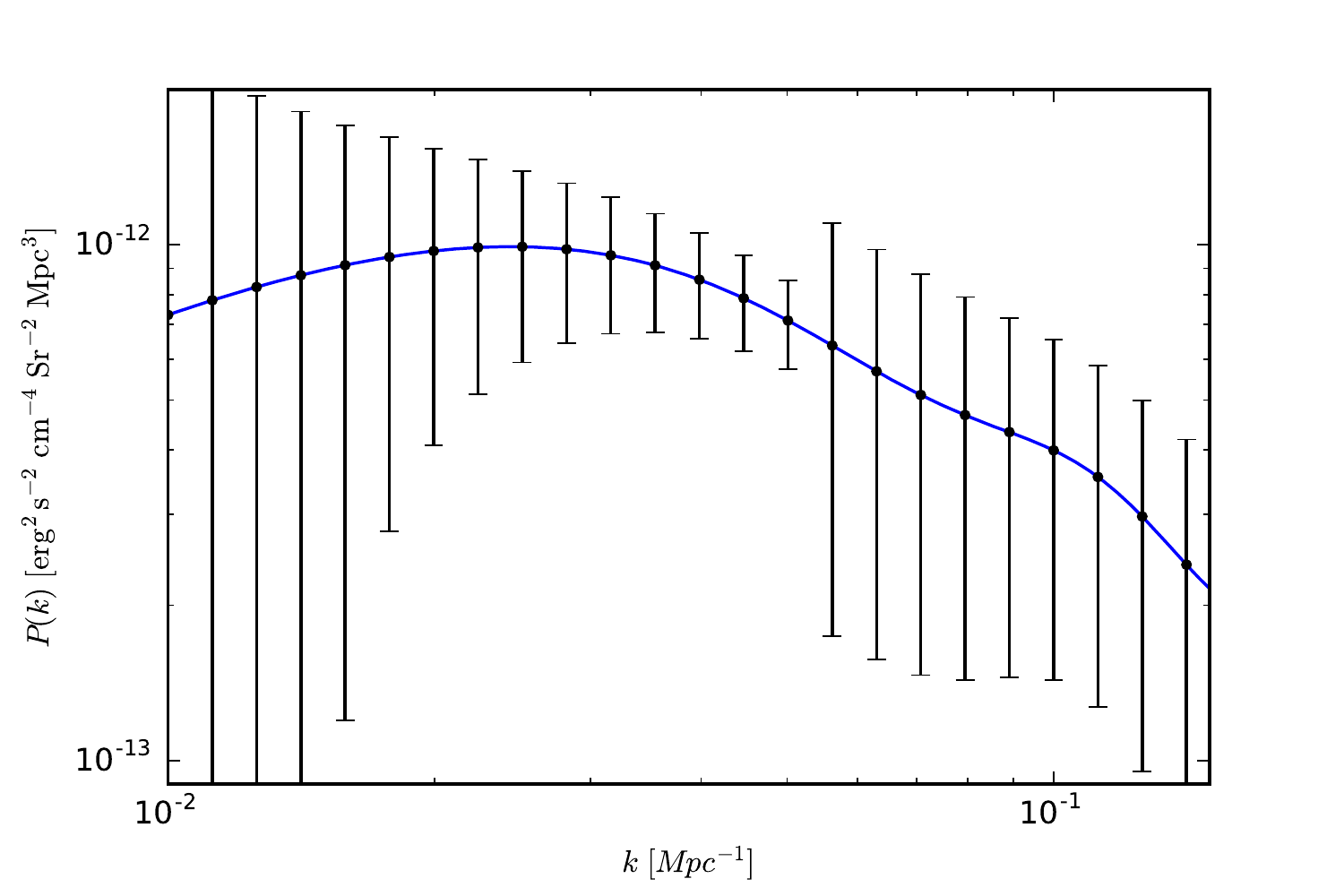}
\caption{OII IM Power Spectrum at $z=1.2$  with forecasted error bar for SPHEREx.}
\label{fig:errpk:oii}
\end{centering}
\end{figure}

OII is mainly an optical and a NIR line and from figure \ref{fig:nuI:comp_opt} we see that there is a short wavelength window where we expect it to be dominant. This broadly corresponds to the transition between optical and NIR. Although one could still use ground telescopes with filters or a spectrograph for OII, IM we will also consider \emph{SPHEREx} \citep{Dore:2014cca} for this line. We take the same instrumental settings as the ones described for $\ha$, but adapt the sensitivity to the required wavelength. For OII at $z=1.2$ we will take $\sigma\l\nu I_\nu\r\sim 3\times 10^{-6}$ erg/s/cm$^2$/Sr. Since we are looking at a different range in the spectrum the redshift resolution shifts to $\delta z=0.05$. We will assume that the sample has $\Delta z=0.4$. Since we are intrinsically looking at lower redshifts, the voxel is smaller in comparison with $\ha$. We therefore have less modes as one can see in figure \ref{fig:pk:oii}. Similarly, instrumental noise is dominant over the power spectrum. With this setting one finds that it will be hard to measure the OII 3D power spectrum using lM. We present the results in figure \ref{fig:errpk:oii}, where it is clear that due to the low volume of the voxel (in comparison with previous lines) we cannot measure the BAO wiggles. There are not enough $k$-modes to overcome the instrumental noise. 

OII thus requires more sensitive experiments with improved angular resolution. This could, in principle, be done by optical ground telescopes equipped with IFUs. To our knowledge no such telescope exists in this wavelength range. As pointed out previously at the end of subsection \ref{subsec:ha}, the proposed Cosmic Dawn Intensity Mapper \citep{Cooray:2016hro} has an improved flux sensitivity that can go from 30 to 50 times higher than SPHEREx. With such improvement we naively expect that the error bars in figure \ref{fig:errpk:oii} to shrink by the square of the same factor.

\subsection{CII IM}

We saw in section \ref{subsec:domline} that approximately above redshift 0.6, CII becomes the dominant line in the FIR and submilimeter part of the spectrum. In this regime one requires radio dishes or telescopes equipped with bolometers at very high altitudes to overcome absorption from the atmosphere.

\begin{table}
\centering            
\caption{Experimental details of ALMA.}
\begin{tabular}{c c c c}        
\hline\hline                 
  Band   & Wavelength & FoV   & T$_{\rm sys}$\\    
  & range(mm) & diameter (") & K\\
\hline
1& 6.7.-9.6  & 139.7  & 26 \\
3& 2.6-3.6  & 54.2  & 60 \\
7& 0.8-1.1  & 16.9  & 219 \\
8 & 0.6-0.8  & 12.6   & 292\\
  \hline                                  
\end{tabular}
\label{tab:alma}     
\end{table}

\begin{table*}
\centering            
\caption{TIME-like experimental details.}
\begin{tabular}{l c c c c c c c c}        
\hline\hline                 
Line & Dish Size & Band   & $\delta\Omega$ & Survey  & NEFD & FoV & Total& $\delta\nu$\\
 
&[m] & [GHz] & [arcmin$^2$]& Area [$\deg^2$]& $[{\rm erg ~s^{-1} cm^{-2} Hz^{-1}} \sqrt{s}$]& [arcmin$^2$]& Time [h]& [GHz]\\
\hline
CII & 6 & 500-700& $0.4^2$& 100& 20$\times 10^{-26}$& $0.4\times25.6$&2000& 0.4 \\
CO(3-2)& 10 & 100-200&$1^2$& 250& 5$\times 10^{-26}$& $1\times64$&2000 & 0.4\\   
  \hline                                  
\end{tabular}
\label{tab:timelike}     
\end{table*}

A radio instrument looking at these frequencies is ALMA (https://almascience.eso.org). ALMA is located in Chajnantor plateau at 5,000 m above sea level and has been constructed to give insights into the birth of stars and stellar systems. It is primarily an interferometer however, in order to cover the required area one would need to make observations in single dish mode. The lowest redshifts are not accessible since only bands 7 and 8 allow observation in single dish mode. Hence we focus on CII emission from $z=2.8$-$5$. One can find the details for these bands in table \ref{tab:alma}. The temperature RMS in a single pixel is given by 
\be
\sigma_T=\frac{T_{sys}}{\sqrt{2N_d\Delta\nu\Delta t}}\,,
\ee
where $T_{sys}$ is the system temperature, $N_d$ is the number of used dishes, $\Delta\nu$ is the frequency resolution and $\Delta t$ is the integration time. In this regime we can use the Rayleigh-Jeans approximation to estimate the intensity rms as
\be
\sigma_{I_\nu}=\frac{2\nu^2k_{\rm B}}{c^2}\sigma_T\,.
\ee
ALMA has a frequency resolution that can go from 3.8kHz to 25MHz which can be tuned accordingly. We compute the 3D power spectrum at $z=3$ using band 8 and $z=4.5$ using band 7. For both we consider $\Delta z=0.4$ and $\delta z=0.05$. Note that we would have to bundle several low resolution bins together for this redshift binning. Due to ALMA constraints, only 4 antennas can work in single dish mode together. We assume 1h of integration per pointing. Unfortunately ALMA cannot do these kind of studies since the errors are several orders of magnitude higher than the power spectrum itself. One would need to lower the system temperature and add more antennas working in single dish mode. 

\begin{figure}
\begin{centering}  
\includegraphics[angle=0,width=\columnwidth]{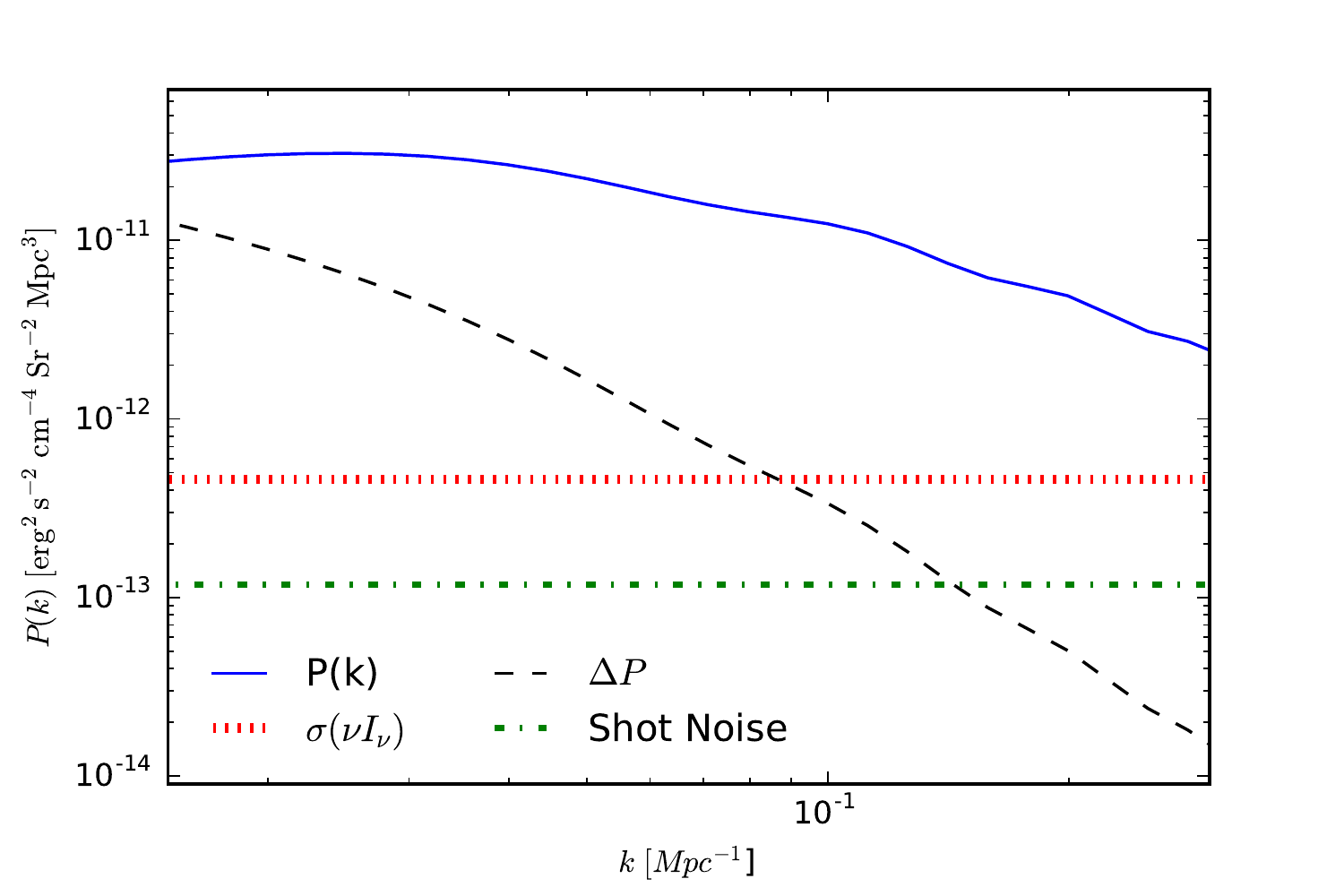}
\caption{Estimated power for CII IM (solid blue), TIME-like experiment noise (thick dotted red), CII shot noise (dashed-dotted green) and the error in estimating the power spectrum (dashed black) at $z=2.2$.}
\label{fig:pk:timecii}
\end{centering}
\end{figure}

\begin{figure}
\begin{centering}  
\includegraphics[angle=0,width=\columnwidth]{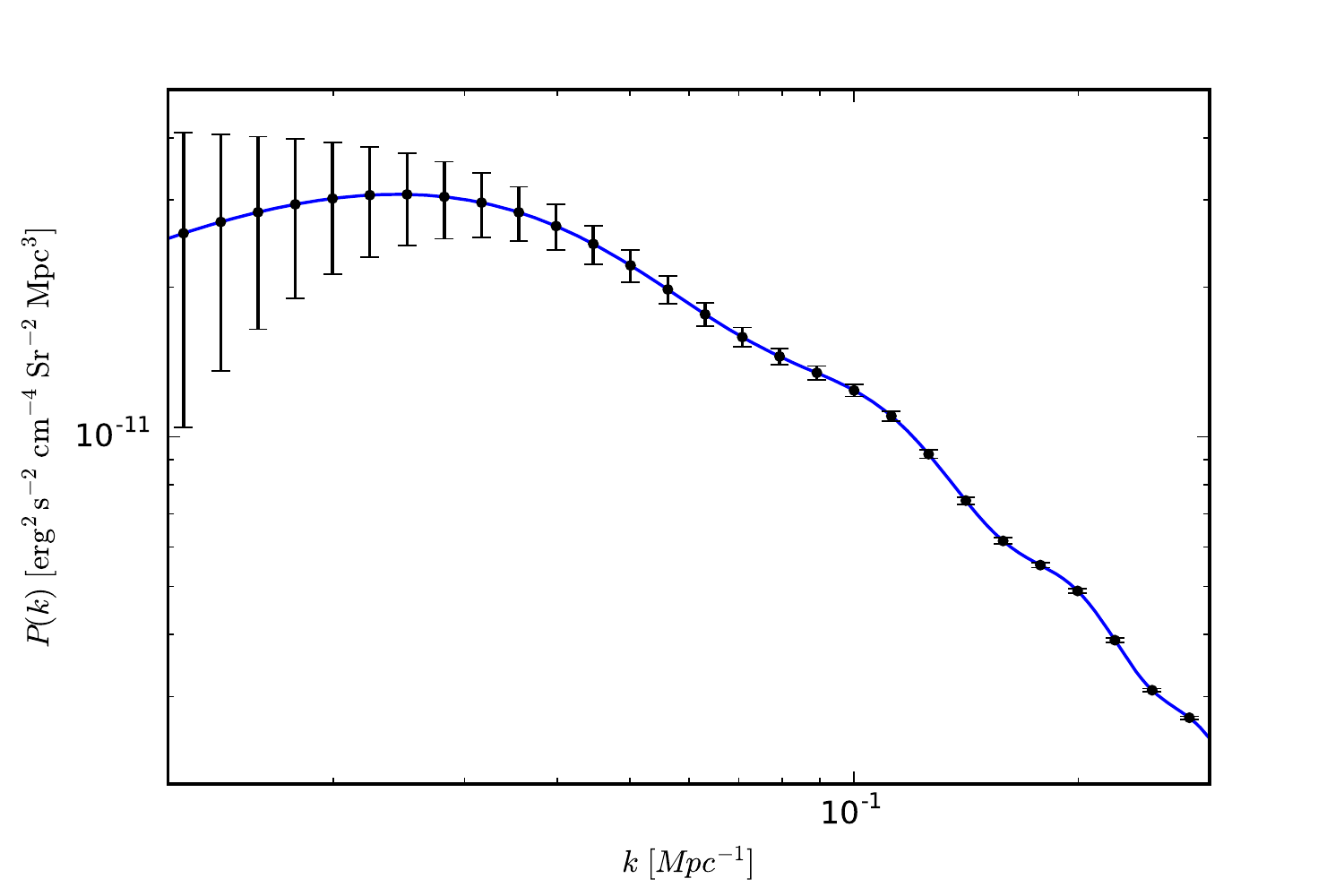}
\caption{CII IM Power Spectrum at $z=2.2$ with forecasted error bar for a TIME-like experiment.}
\label{fig:errpk:timecii}
\end{centering}
\end{figure}
 
Another possibility is to consider a CII IM experiment inspired by TIME \citep{2014JLTP..176..767S}, which is a proposed CII intensity mapping for EoR. Since we want to study the late universe with IM experiments we need to adapt the dish size and angular resolution for the targeted line. In table \ref{tab:timelike} we summarize the assumed experimental details. For full details of TIME and TIME-\emph{Pilot} please refer to \citet{2014JLTP..176..767S} and to \citet{2014SPIE.9153E..1WC}. Note that we assume the number of bolometers is the same. We also considered a higher noise since we expect the noise to increase with frequency. For this experiment we decided to target emission from a central redshift of $z=2.2$ and assumed a sample with $\Delta z=0.4$ and a survey of 100 $\deg^2$. This is already enough for the instrumental noise power spectrum to go well below the expected power spectrum, as we can see in figure \ref{fig:pk:timecii}. This would be as good as HETDEX for $\lya$ but looking at a complementary tracer of the dark matter distribution. The low noise level becomes clear when we plot the 3D power spectrum with forecasted error bars in figure \ref{fig:errpk:timecii}. One should note that these IM experiments have low angular resolution, but it is still enough to see the BAO wiggles in the power spectrum. We do not show a zoomed in part of the power spectrum because the TIME-like CII IM experiment is merely a proposal, in opposition to HETDEX (ongoing) and \emph{SPHEREx} (to be approved). Still we want to stress that a CII experiment with this configuration could measure the BAOs and give a complementary probe of the large-scale distribution of matter in the universe.

We have not exhausted the range of possible experiments for CII IM. One should still refer to the work of \citet{Uzgil:2014pga}, which considers using an atmospheric balloon or a cryogenic satellite to perform CII IM. Satellite experiments have the clear advantage of not suffering from FIR absorption from moisture. Still, ground experiments at high altitudes can largely minimize such issues, as ALMA does. 

\subsection{CO}

We saw in section \ref{subsec:domline} that the three lowest CO rotational lines are the best candidates to be used as IM tracers of the underlying dark matter distribution. Currently COPPS \citep{Keating:2016pka} is looking for the CO(1-0) cosmological signal to recover the power spectrum. Similarly COMAP \citep{Li:2015gqa} will be looking at a window from 30GHz to 34GHz, hence looking at CO(1-0) emission from $z=2.4-2.8$ and CO(2-1) emission at late EoR redshifts, $z = 5.8-6.7$. Although it will have high spectral resolution it will cover a very limited area of the sky. Since these experiments already exists with the specific goal of doing IM of CO(1-0) we will focus on other CO rotation lines. Another reason not to look at this line is because synchrotron emission from galaxies and jets from quasars peak at these frequencies. Therefore let us focus, for now, on the other two CO rotational lines, CO(3-2) and CO(2-1). Bands 4 and 1 of ALMA (see table \ref{tab:alma}) will look at these frequencies, respectively. Unfortunately, ALMA performs as badly as for CO as it does for CII IM.

\begin{figure}
\begin{centering}  
\includegraphics[angle=0,width=\columnwidth]{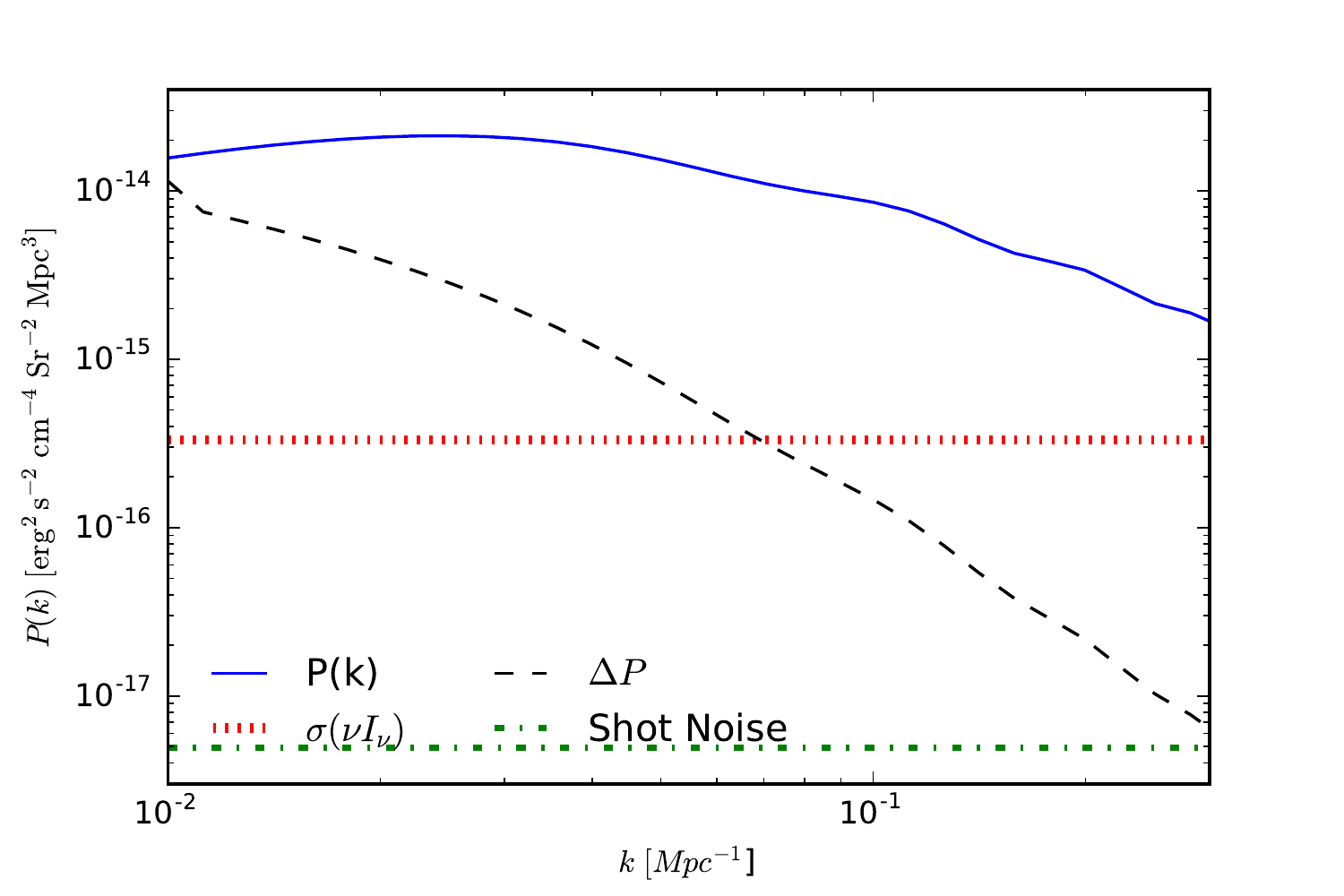}
\caption{Estimated power for CO(3-2) IM (solid blue), TIME-like experiment noise (thick dotted red), CO(3-2) shot noise (dashed-dotted green) and the error in estimating the power spectrum (dashed black) at $z=2.0$.}
\label{fig:pk:timeco}
\end{centering}
\end{figure}

\begin{figure}
\begin{centering}  
\includegraphics[angle=0,width=\columnwidth]{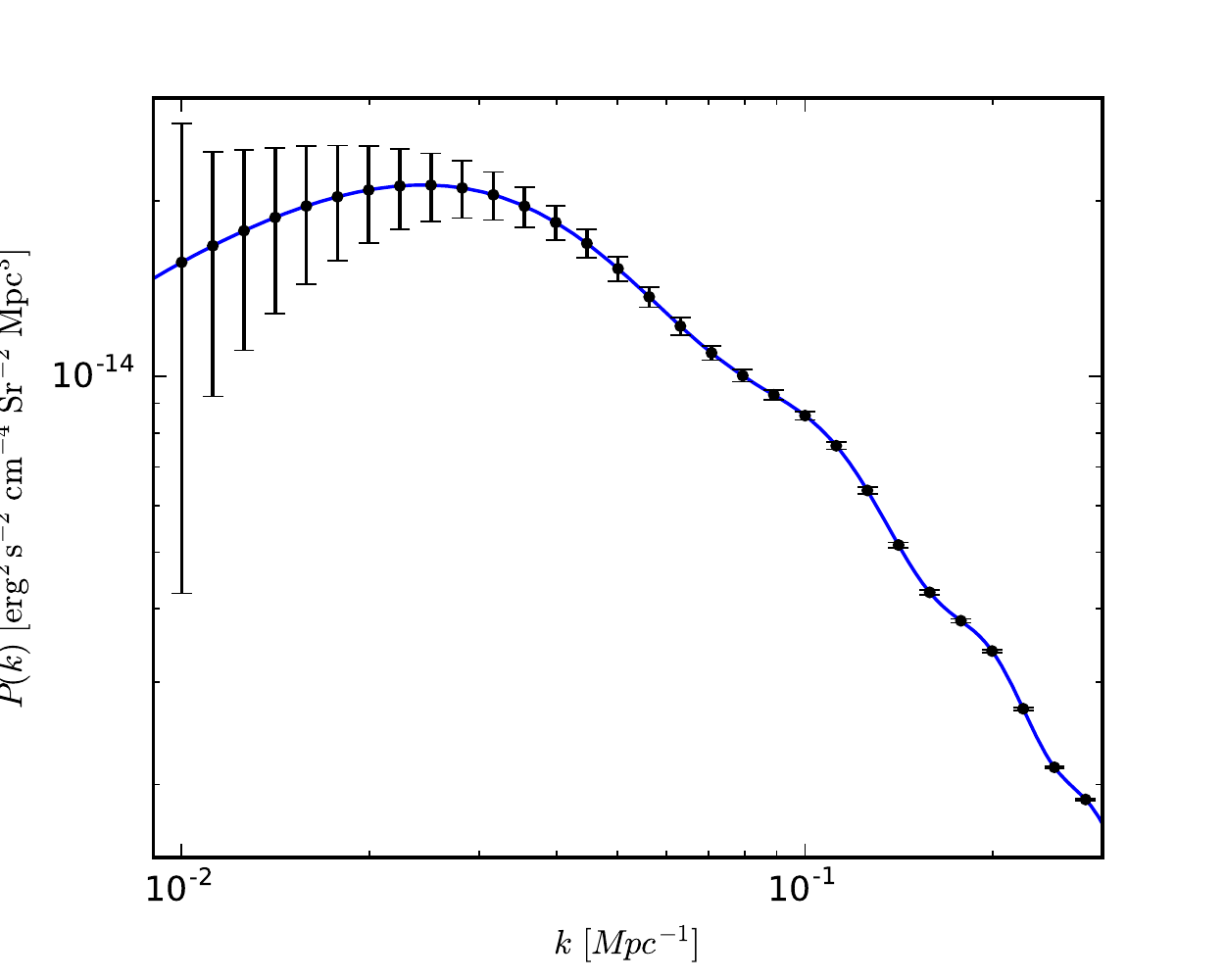}
\caption{CO(3-2) IM Power Spectrum at $z=2.0$ with forecasted error bar for a TIME-like experiment.}
\label{fig:errpk:timeco}
\end{centering}
\end{figure}

The last experimental set up that we consider is again a TIME-like experiment to perform CO(3-2) IM.  We change the experimental setting to a 10m dish which will give around 1' angular resolution, as one can see in table \ref{tab:timelike}. This lower resolution allows for a larger survey area for the same survey time, which we take to be 250 $\deg^2$. As a test example we consider a sample survey with $\Delta z =0.4$ around a central redshift of $z=2.0$ and with a redshift resolution of $\delta z=0.01$. Figure \ref{fig:pk:timeco} shows the power spectrum of CO(3-2) IM (solid blue), the power coming from the TIME-like instrumental noise (dotted red), CO(3-2) shot noise (dot-dashed green) and $\Delta P_{\rm CO(3-2)}$ (dashed black) at $z=2.0$. Although the instrumental noise is higher than the 3D power spectrum there are enough $k$-modes such that $\Delta P$ falls below the power spectrum itself. This becomes clearer in figure \ref{fig:errpk:timeco} where we can see that one could even measure the BAOs.

\section{IGM emission} \label{sec:IGM}
In the circumgalactic medium (CGM), the gas is warm/hot and at low redshifts one expects it to be considerably metal rich. This medium can be observed because of its far-UV and X-ray emission lines. However, the strength of these emission lines is considerably smaller than ISM emission in the same part of the spectrum. Also, given the large pixels involved in IM studies the power spectra from CGM emission follows that of galaxies and so it has little impact for the proposed large scale studies.
In the colder IGM filaments the composition of the gas is close to primordial. Therefore, its main radiative cooling channels should be emission in the $\lya$ and the $\ha$ lines.
The intensity of these lines from recombinations in IGM filaments should be of the order of 10\% of the emission from galaxies and its power spectra should be relatively flat compared to the emission from galaxies in the same lines \citep{Silva:2012mtb,2016arXiv160306952S}.
Given the large cross section of photons of the Lyman series, there will be absorption of UV background photons (mainly between the Lyman $\beta$ and Lyman $\alpha$ frequencies) in the neutral hydrogen gas contained in the cold filaments. These photons will be reemitted as $\lya$ photons which follow the spatial distribution of the gas filaments. The importance of this emission depends on the average density of the hydrogen gas in the filaments, which includes gas in clouds with lower column densities than the gas probed by the Lyman-alpha forest. It also depends in the intensity of the UV background in the relevant frequency range and at the relevant redshifts, which is only known within one order of magnitude at most. Consistently, modeling the intensity of the scattered radiation requires detailed simulations and the use of radiative transfer codes, which goes beyond the scope of this study. The intensity of this emission should however be smaller than the emission from galaxies. Also, the smaller bias of IGM emission and the flatness of its spectra will, in any case, make it subdominant relative to the galaxies' power spectra in the post Reionization Epoch.

\begin{table*}
\centering            
\caption{Survey details in estimating $P(k)$ for the different lines in consideration.}
\begin{tabular}{l c c c c c c c c c}        
\hline\hline                 
 & Line & Area [$\deg^2$] & z  & $\delta z$ & $\delta \Omega$ [$10^{-9}$ Sr]& $\Delta z$ & $P_N$ [erg$^2$s$^{-2}$cm$^{-4}$Sr$^{-2}$Mpc$^3$] &$k$-range [Mpc$^{-1}$]& SNR \\
%  &  &  [$\deg^2$] &   &  &  [$10^{-9}$ Sr]&  &  [erg$^2$s$^{-2}$cm$^{-4}$Sr$^{-2}$Mpc$^3$] & [Mpc$^{-1}$]&  \\
\hline
Hetdex& $\lya$& 300& 2.1 & 0.005 & $0.213$& 0.4 &7.24 $\times 10^{-16}$&0.009-0.3& 489\\
SPHEREx& $\ha $& 7000 & 1.9 & 0.07  & $0.903$& 0.4 & 2.59 $\times 10^{-12}$&0.009-0.3& 105 \\
&  OII& 7000 &  1.2 & 0.05 & $0.903$& 0.4 &  1.34 $\times 10^{-11}$ & 0.02-0.1& 11\\

%CCAT-like& CII $\&$ CO(3-2)&  [$\deg^2$]& & [Sr]&\\
TIME-like& CII &  100  & 2.2 &0.002 & $13.5$& 0.4& 4.59 $\times 10^{-13}$ & 0.01-0.3& 294\\
& &  2 (50h) &  &&&& 3.67 $\times 10^{-13}$  & 0.05-0.3& 42\\

& CO(3-2) &   250 & 2.0 &0.01 & $85$& 0.4& 3.31 $\times 10^{-16}$ & 0.01-0.3 & 471 \\
&  &   2 (50h) & & & & & 1.06 $\times 10^{-16}$ & 0.05-0.3 & 45 \\

  \hline                                  
\end{tabular}
\label{tab:expsum}     
\end{table*}

\section{Contamination in IM} \label{sec:Foregrounds}

This study indicates the best lines/redshifts for performing IM studies, assuming that the foregrounds of each line could be efficiently removed. Unfortunately, this is not always the case and so a prior detailed study of a line's foregrounds/backgrounds contamination is required. Only then one can seriously use these lines as cosmological probes. The contamination study should be made for a specific target line and take into account the proposed experimental setup, the available foreground removal methods, the target spatial range in the power spectra and the allowed error for the proposed cosmological/astrophysical goals.

This indeed sets the way for future work in this field. To properly access the feasibility of all these cleaning techniques one has to use simulations of the 3D distribution of halos. For each halo in a line-of-sight one attributes a luminosity from each line (including all subdominant lines) as prescribed in section \ref{sec:lineemission}, as well as a continuum emission. Similarly one should assume a continuous extragalactic background. Then, by redshifting all the emission to the observers frame, one should access how well one can disentangle the signal assuming background smoothness and cross-correlations. This possibility not only opens new windows to probe the distribution of matter in the universe, but also to study global properties of galaxies and their astrophysical properties. Similarly, different lines will be present in different halo ranges and redshift ranges. These studies open the possibility to use IM to study the chemical evolution of the universe as well as the halo mass function.

\subsection{Line contamination}

IM will suffer contamination from other lines in the foreground or background of the target sources, which are observed at the same frequency range. Given that in IM studies sources are not resolved, one cannot easily differentiate between the different emission lines. Contamination from interloping lines will show up in the observed power spectrum projected into the redshift of the target line. The power spectra of foreground lines will have a higher contamination with respect to what is suggested by figures \ref{fig:nuI:comp_opt}, \ref{fig:nuI:comp_fir} and \ref{fig:nuI:comp_cos} \citep{Gong:2013xda}. As a result, even in the cases when the intensity of the target line is dominating the signal, it will usually be necessary to use a cleaning method to deal with the bright line foregrounds.
It has however been shown, in several previous studies, that for an intensity mapping experiment with a resolution similar to the proposed experiments described in Section \ref{sec:Surveys}, the required masking of bright sources will only cover a small percentage of the pixels in the observational maps.
The success of this masking procedure is due to the contamination being highly dominated by a small number of bright foreground sources and so it can be reduced by a few orders of magnitude with this procedure. Also, the signal and the foregrounds positions are uncorrelated and so for small masking percentages the power of the target signal will practically not be attenuated.

Nonetheless an efficient method of dealing with line foregrounds/backgrounds is to cross-correlate the target line with the galaxies number density or with other line emission at the same redshift. This can in principle be used to recover the power spectrum of the target line free of contamination at a first order approximation. This technique is not only powerful to clean contaminants but also to increase the significance of the detection of over-densities of emission. 

\subsection{Continuum background and foreground contamination}

Additional contamination will come from the extragalactic continuum background. This background originates from AGN emission and continuum stellar, free-free, free-bound and two photon emission from star forming galaxies and it can be estimated as in \cite{2012ApJ...756...92C}. This radiation can be 
fitted out from intensity maps due to its smooth evolution in frequency compared to the line intensity fluctuations. 
Besides its smoothness, one can use cross-correlations between different emission lines, which suffer contamination from different parts of the spectrum, to at first order obtain a clean signal.
Although this is generically true, the background contamination will always be present in the power spectrum error determination. Hence, one can immediately foresee that large areas with enough $k$-modes are needed to reduce cross-correlation errors. 

Continuum foregrounds in intensity maps, of the UV and optical lines studied here, are not as well known as the continuum foregrounds in the radio band. These have been intensely studied in the context of 21cm IM. However, the continuum contamination relative to the intensity of the signal from the target line should be much smaller for UV and optical lines than that for radio lines (even if it is still dominating the signal) and so the residuals from the removal of these contaminants are less problematic.

\section{Discussion} \label{sec:discussion}

IM studies open new windows to probe not only the distribution of matter in the universe, but also to study global properties of galaxies. Particularly IM is useful to probe the BAO scale independently from galaxy surveys. The evolution of the halo mass function is also an interesting cosmological goal although it will be difficult to disentangle it from the evolution of astrophysical properties of galaxies. On the astrophysical side IM of metal lines can be used to probe the poorly known chemical evolution of the gas in the universe.

In table \ref{tab:expsum} we summarize the experimental and sample details for each line we consider for IM. In the last two columns we present the signal-to-noise ratio for each survey for a fixed $k$-range. Although higher SNRs can be attained going to smaller scales, we caped at $k=0.3$ Mpc$^{-1}$ so we do not need to worry about issues with non-linear scales. Also, below these scales the voxel of IM may be too small for the assumption that astrophysical fluctuations between galaxies average out. From the last column of table \ref{tab:expsum} we see that using HETDEX as a $\lya$ IM survey would be the best performing survey out of the ones considered in this paper. Although $\ha$ IM with SPHEREx is possible, its SNR is not as good as for $\lya$ or the FIR lines with a TIME-like experiment. 
The proposed OII survey will have the worst performance of the IM experiments considered, given the smaller physical volume to be observed. This is due to the relatively low redshifts of the observational frequency window where OII is the dominant emission line. CII with a TIME-like experiment also has a high SNR but it is still lower than CO(3-2). This is mainly due to the fact that for a fixed survey time a CO(3-2) IM survey can sample a bigger area of the sky. In fact, a CO(3-2) IM survey can do as well as $\lya$ with HETDEX. Since TIME-like experiments are in the realm of possibilities we also performed the calculations for a pilot experiment. For both CII and CO(3-2) we computed the SNR for a smaller 50h survey covering 2$\deg^2$ of the sky. These can be used as proof of concept despite the lower SNR. Looking at table \ref{tab:expsum} one can say that $\lya$ and CO(3-2) would be the best lines to perform IM. This is not necessarily true due to contamination issues. As said before, different lines are sensitive to different mass ranges of the HMF, to the chemical evolution of the Universe and to the underlying astrophysical processes within galaxies. Hence, we would like to stress that all lines should be seen as complementary, although $\lya$ is the best starting point. 

This work is intended to probe the new windows opened by line IM for studying our cosmology. On top of looking at the distribution of matter in the Universe, one can study the astrophysical properties of galaxies. We started by modelling the luminosity of a line in terms of halo mass. We proposed a general prescription based on previous theoretical and observational studies. Although one expects scatter in these relations on small scales it should average out in big enough voxels. Under our models we find figures \ref{fig:nuI:comp_opt}, \ref{fig:nuI:comp_fir} and \ref{fig:nuI:comp_cos} which indicate which are the best candidates for line IM. Unsurprisingly, these are $\lya$, OII, $\ha$, CII and the lowest rotation lines of CO. We then follow by estimating power spectrum error bars for each line with reasonable experimental settings for a fixed redshift range. We find that one can measure the power spectrum of these lines, assuming a cleaned signal. %This last assumption is crucial to perform IM studies. 
Unlike 21cm emission of HI, UV, optical and infrared lines have stronger line contaminants that need to be considered. We discussed foreground contamination and ways to extract the target signal from observational maps in Section \ref{sec:Foregrounds} and found that there are promising ways to successfully clean many of these maps. The literature already has a wide discussion on such methods either for the EoR \citep{Visbal:2010rz,Lidz:2011dx,Silva:2012mtb,Gong:2013xda,Silva:2014ira,Breysse:2015baa} or for the late universe \citep{Croft:2015nna,Lidz:2016lub} which we will explore in the future.

This work considers the 3D power spectrum P(k), as it is conventional in galaxy surveys. Nonetheless, one should point out that IM allows for tomographic studies using the angular power spectra $C_\ell$. There are several advantages on doing so. First of all P(k) is gauge dependent and suffers from projection effects, which does not happen when using $C_\ell$. Furthermore, the contributions from lensing and the so called GR corrections to number count fluctuations \citep{Yoo:2010ni,Challinor:2011bk,Bonvin:2011bg} and temperature/intensity fluctuations \citep{Hall:2012wd} are easily included in the angular power but not in the 3D P(k).   

We note that in order to successfully do science with line intensity maps it will be necessary to perform further detailed studies.  
Galaxy emission studies will have to be done for foreground/interloping lines. Also, continuum contamination estimates should be made accounting for dust absorption and/or propagation in the IGM. 
%Galaxy emission will have to include unwanted lines and continuum emission contamination on top of the targeted line. 
This is needed to properly access the feasibility of all contamination cleaning techniques. As prescribed in section \ref{sec:lineemission} one should use cosmological simulations of the 3D distribution of halos and attribute line luminosities and  galactic continuum emission to each on top of a continuous extragalactic continuum. Then, by redshifting all the emission to the observers' frame, one should assess how well one can disentangle the signal, accounting for masking rates of interloping lines, the frequency smoothness of the continuum contamination and cross-correlations. %Unfortunately there is no general way of preceding and this has to be done targeted line by line.   
These are general guidelines since the ideal method depends on the target line.

We conclude by emphasizing that IM, with lines other than HI, shows great potential to measure the large-scale distribution of matter in the Universe. This is especially true since independent sets of lines can measure the BAO scale around $z=2$, a further test of the expansion of the universe at intermediate redshifts. One should also note that these lines acquire an important role in the context of the multi-tracer technique \citep{Seljak:2008xr,McDonald:2008sh} to beat cosmic variance. As \citet{2015ApJ...812L..22F} showed, the improvement in measuring large scale effects is greatly increased as the ratio of the two biases deviates from $\sim1$. Hence, these lines can give a better bias ratio for a particular paring of DM tracers. 

\subsection*{Acknowledgments}
We would like to thank Eiichiro Komatsu for useful discussions. We thank Imogen Whittam for proof reading the paper.
J. F. and M.G.S. acknowledges support from South African Square Kilometre Array Project and National Research Foundation. M.B.S thanks the Netherlands Foundation for Scientific Research support through the VICI grant 639.043.006. M.G.S. also thanks the support of FCT under the grant PTDC/FIS-AST/2194/2012. A.C. acknowledges support from the NSF grant AST-1313319 and the NASA grants NNX16AF39G and NNX16AF38G.

\bibliographystyle{mn2e}

\label{lastpage}

\end{document}